\documentclass[draft,dvipsnames,table]{agujournal2019}
\usepackage{url} 
\usepackage{lineno}
\usepackage[inline]{trackchanges} 
\usepackage{soul}
\usepackage{amsmath}

\definecolor{CellGreen}{HTML}{DDFFE0}
\definecolor{CellRed}{HTML}{FFDDDD}

\def\datastatement{\vskip12pt\noindent{\bf
Data Availability Statement\vrule depth 6pt
width0pt\relax}\\*\noindent\ignorespaces}
\draftfalse

\journalname{Journal of Advances in Modeling Earth Systems}

\def\deg{$^{\circ}\,$}

\begin{document}

%
%

\title{Sub-seasonal forecasting with a large ensemble of deep-learning weather prediction models}

%
%




\authors{Jonathan A. Weyn\affil{1}\thanks{Current affiliation: Microsoft, Redmond, WA, USA.}, Dale R. Durran\affil{1}, Rich Caruana\affil{2}, Nathaniel Cresswell-Clay\affil{1}}


\affiliation{1}{Department of Atmospheric Sciences, University of Washington}
\affiliation{2}{Microsoft Research, Redmond, Washington}




\correspondingauthor{J. Weyn}{jweyn@uw.edu}




\begin{keypoints}
\item An ensemble forecast system is developed using convolution neural networks (CNNs) to generate data-driven global forecasts.
\item Only 3 seconds are required to compute a large 320-member ensemble of skillful 6-week sub-seasonal predictions.
\item Shorter lead time forecasts also show skill, including a single deterministic 4-day forecast for Hurricane Irma.
\end{keypoints}

%
%

%
%


\begin{abstract}
We present an ensemble prediction system using a Deep Learning Weather Prediction (DLWP) model that recursively predicts  key atmospheric variables with six-hour time resolution.  This model uses convolutional neural networks (CNNs) on a cubed sphere grid to produce global forecasts.  The approach is computationally efficient, requiring just three minutes on a single GPU to produce a 320-member set of six-week forecasts at 1.4\deg resolution. Ensemble spread is primarily produced by randomizing the CNN training process to create a set of 32 DLWP models with slightly different learned weights. 

Although our DLWP model does not forecast precipitation, it does forecast total column water vapor, and it gives a reasonable 4.5-day deterministic forecast of Hurricane Irma. In addition to simulating mid-latitude weather systems, it spontaneously generates tropical cyclones in a one-year free-running simulation.  Averaged globally and over a two-year test set, the ensemble mean RMSE retains skill relative to climatology beyond two-weeks, with anomaly correlation coefficients remaining above 0.6 through six days.

Our primary application is to subseasonal-to-seasonal (S2S) forecasting at lead times from two to six weeks. Current forecast systems have low skill in predicting one- or 2-week-average weather patterns at S2S time scales. The continuous ranked probability score (CRPS) and the ranked probability skill score (RPSS) show that the DLWP ensemble is only modestly inferior in performance to the European Centre for Medium Range Weather Forecasts (ECMWF) S2S ensemble over land at lead times of 4 and 5-6 weeks. At shorter lead times, the ECMWF ensemble performs better than DLWP.

\end{abstract}

\section*{Plain Language Summary}
We develop a machine learning weather prediction system, trained on past weather data, thereby providing an alternative approach to the current widely used computer models that predict the weather based on approximate mathematical representations of physical laws.  Our approach is much more computationally efficient, allowing us to make a very large number of similar forecasts, known as an ensemble, for the same weather event.  The ensemble better defines the range of possible future weather conditions than a single forecast.  Our ensembles show skill at both short forecast lead times and for 4-6-week (sub-seasonal) prediction.  The sub-seasonal predictions are for 1 or 2-week averaged weather patterns. Although our forecasts are better than benchmark guesses, such as a forecast consisting of the average weather conditions on a given calendar date, the quantitative forecast skill remains low for sub-seasonal prediction.  Nevertheless, by several quantitative measures our forecasts at 4-6 weeks score nearly as well as the current state-of-the-art forecasts from major operational centers.  The ease with which our machine learning approach can efficiently generate very large ensemble forecasts holds promise for future developments to improve the skill of sub-seasonal prediction.

%
%

%


%
%
%
%

\section{Introduction}

Weather forecasting relies heavily on data assimilation to estimate the current state of the atmosphere and on numerical weather prediction (NWP) to approximate its subsequent evolution. The skill of such  deterministic weather forecasts is typically limited to about two weeks by the chaotic growth of small initial errors and inaccuracies in our approximate models of the atmosphere. On much longer, multi-month time scales, the coupling of the atmosphere with slowly evolving ocean-land forcing allows skillful seasonal forecasts of monthly or seasonally averaged conditions. Between these two extremes, the production of skillful one- or two-week averaged forecasts at lead times ranging roughly between two weeks and two months (the so-called subseasonal-to-seasonal or S2S time frame) has proved particularly challenging; yet there are many societal sectors that would greatly benefit from improved S2S forecasts \cite{White2017}.  Several major operational centers have developed NWP-based ensemble systems focused on improving S2S forecasting \cite{Vitart2017}.

In 1992, the European Centre for Medium-Range Weather Forecasts (ECMWF) and the National Centers for Environmental Prediction (NCEP) began issuing ensemble weather forecasts. They were soon followed by the other major weather prediction centers across the world. 
Ensemble forecasts strive to provide a set of equally-likely forecast realizations spanning the range of possible future atmospheric states.  
Early evidence for the economic value of probabilistic forecasts derived from the ECMWF ensemble relative to a single deterministic forecast was provided by \citeA{Richardson2000}. Ensemble forecasts are now recognized as essential to represent the probabilistic nature of weather forecasting and to break through the intrinsic limits to predictability of the atmosphere \cite{Palmer2018}.

Ensembles are particularly appropriate as one looks beyond lead times where deterministic forecasts lose all skill relative to climatology. On S2S lead times, ensemble-mean and ensemble-based probabilistic forecasts have shown modest skill relative to climatology \cite{Vitart2004, Weigel2008,Vitart2014,Monhart2018}.
Computational resources do, however, impose a significant limitation on efforts to create S2S forecasts with NWP ensembles.  As of 2016, 11 forecast centers were contributing S2S forecasts to the S2S database \cite{Vitart2017}, and the ensembles from three of these  
centers consisted of just four members.  The largest S2S ensemble, with 51 members providing forecasts out to 46 days, is generated by ECMWF's  Integrated Forecast System (IFS) using a significant fraction of the time available on one of the world's most powerful computer systems \cite{Bauer2015}. Yet there is increasing evidence that the number of ensemble members for S2S forecasts should be higher, perhaps in the 100-200 range \cite{Buizza2019}. Large ensemble sizes are also helpful for assessing the likelihood of events in the tails of probabilistic forecast distributions \cite{Leutbecher2018}, and such extreme events are often the most impactful.

Machine learning provides one potential avenue to develop S2S forecasts systems with significantly lower computational costs. Recognizing that there are other successful machine-learning approaches to S2S forecasting \cite{Hwang2019}, here our focus will be on the development of a data-driven deep-learning weather prediction (DLWP) model that can be iteratively stepped forward, like traditional NWP models, to simulate atmospheric states at arbitrarily long lead times. In one of the first attempts to use ML to create such a model,  \citeA{Dueben2018} trained neural networks (NNs) on several years of reanalysis data to predict 500~hPa geopotential height ($Z_{500}$) on the globe at 6\deg resolution, demonstrating the ability to produce ML weather forecasts that have at least modest forecast skill. 
Using advanced convolutional neural networks (CNNs), \citeA{Scher2019} trained algorithms on simulations from a simplified GCM that significantly outperformed baseline metrics and effectively captured the simplified-GCM dynamics at spherical harmonic resolutions of T21 and T42 (roughly 5.6\deg and 2.8\deg). Training only on historical data, \citeA{Weyn2019} used CNNs to generate forecasts for northern mid-latitude  $Z_{500}$ and 300-700-hPa thickness ($\tau_{300-700}$) on a 2.5\deg latitude-longitude grid that showed skill relative to climatology and persistence through five days.

Recently we extended our DLWP model to the full globe using a volume-conservative mapping to project global data from latitude-longitude grids onto a cubed sphere and improved the CNN architecture operating on the cube faces \cite[hereafter WDC20]{Weyn2020}.  In addition to $Z_{500}$ and $\tau_{300-700}$,
our improved model forecasts two additional surface fields, 1000 hPa height ($Z_{1000}$) and 2-meter temperature ($T_2$), and uses three externally-specified 2D fields: a land-sea mask, topographic height, and top of the atmosphere insolation. This new 1.9\deg resolution model showed skill relative to climatology and persistence through seven days.  Moreover, it could be stepped forward repeatedly from a single initialization for at least one year, and while doing so, captured the seasonal cycle with reasonable accuracy.

In the following we further improve our DLWP model by adding two more 2D prognostic fields and increasing the spatial resolution to 1.4\deg. Large 320-member ensembles generated using the improved model are used to provide S2S forecasts through a six-week lead time.  These forecasts are verified against ERA5 data and compared to operational ECMWF S2S products.

The remainder of this paper is organized as follows.  Section 2 describes the improvements to our previous DLWP model, while section 3 discusses the incorporation of that model in our ensemble forecast system.  The behavior of that ensemble is assessed at short deterministic lead times in section 4, and at longer S2S lead times in section 5.  Section 6 contains the conclusions.

\section{The DLWP model}

The basic model is very similar to that described in detail in WDC20, in which four forecast fields, geopotential height at 1000 hPa ($Z_{1000}$) and at 500 hPa ($Z_{500}$), 300-700 hPa thickness ($\tau_{300-700}$), and 2-meter temperature ($T_2$), are mapped to a cubed sphere. Three known fields are also provided: top-of-atmosphere radiation, topographic height, and a land-sea mask. Convolutional neural networks (CNN) were trained using the same 3$\times$3 set of horizontal spatial filters on all four equatorial faces on the cube (WDC20, Fig.~1). A different set of filters was applied to the two polar faces.  A U-Net architecture \cite{Ronneberger2015} with skip connections is employed to capture multi-scale processes via average pooling and corresponding up-sampling.  The skip connections across each level of spatial refinement ensure high-resolution information is preserved.  The activation functions are leaky ReLU functions capped at a scaled value of 10. The model is recursively stepped forward with 12-hr time steps, such that a single step maps the fields at two time levels $t_0-6$ and $t_0$ hr to forecast fields at $t_0+6$ and $t_0+12$ hr.  The model is trained to minimize the mean squared error (MSE) over two steps, or equivalently over a 24-hr period with 6-hr temporal resolution.

Here we extend the WDC20 model by adding two more forecast fields: temperature at 850 hPa ($T_{850}$), which is strongly modulated by  large-scale weather patterns while exhibiting less sensitivity to diurnal heating and surface-layer processes than $T_2$, and total column water vapor (TCWV). TCWV is the vertically-integrated total gas-phase water above each grid cell; its inclusion is a step toward characterizing tropical convective systems including tropical cyclones and  the Madden-Julian oscillation (MJO).  The MJO is believed to be an important source of forecast skill on S2S time scales and has been characterized as a ``moisture mode" \cite{adames2016}.

The resolution was also modestly increased from $48\times 48$ to  $64\times 64$ grid cells on each face of the cube, yielding an effective resolution of approximately 1.4$^{\circ}$ in latitude and longitude at the equator.
ERA5 data at a gridded resolution of 1$^{\circ}$ in latitude and longitude were remapped with the Tempest-Remap package \cite{Ullrich2015,Ullrich2016} for training, validation and testing.  Somewhat fortuitously, the WDC20 convolutional neural network architecture continued to perform quite well despite the changes to the model input and target data, although  we were able to improve the model by doubling the number of filters used in each convolutional layer.  This increased the number of filters in the first layer from 32 to 64. 
Our improved DLWP CNN architecture is tabulated in Table~\ref{unet-table-64}. The increases in the number of filters in each layer and number of forecast fields increased the total number of trainable parameters relative to that in WDC20 by about a factor of about 4, to 2.7 million. Nevertheless, as a result of additional code optimization, the model still trains in 6--8 days.

\begin{table}[t]
    \caption{\label{unet-table-64}  \small CNN architecture for DLWP as a sequence of operations on layers. The parameter $v$ represents the number of input fields, $t$ represents the number of input time steps, and $c$ represents the number of auxiliary prescribed inputs (here top-of-atmosphere radiation at $t$ times, land-sea mask and topographic height). The layer names (except for the suffix ``CubeSphere") correspond to the names in the Keras Library. ``Concatenate" appends the state in parentheses, numbered earlier, to the output of the previous layer.}
    \begin{center}
    \begin{tabular}{ccccc}
    \hline
    Layer & Filters & Filter size & Output shape$^a$ & Trainable params$^b$ \\
    \hline
    $input$ & & & (6, 64, 64, $vt + c$) & \\
    \hline
    Conv2D--CubeSphere & 64 & $3\times3$ & (6, 64, 64, 64) & 18,560 \\
    Conv2D--CubeSphere (1) & 64 & $3\times3$ & (6, 64, 64, 64) & 73,856\\
    AveragePooling2D & & $2\times2$ & (6, 32, 32, 64) & \\
    Conv2D--CubeSphere & 128 & $3\times3$ & (6, 32, 32, 128) & 147,712 \\
    Conv2D--CubeSphere (2) & 128 & $3\times3$ & (6, 32, 32, 128) & 295,168 \\
    AveragePooling2D & & $2\times2$ & (6, 16, 16, 128) & \\
    Conv2D--CubeSphere & 256 & $3\times3$ & (6, 16, 16, 256) & 590,336 \\
    Conv2D--CubeSphere & 128 & $3\times3$ & (6, 16, 16, 128) & 590,080 \\
    UpSampling2D & & $2\times2$ & (6, 32, 32, 128) & \\
    Concatenate (2) & & & (6, 32, 32, 256) & \\
    Conv2D--CubeSphere & 128 & $3\times3$ & (6, 32, 32, 128) & 590,080 \\
    Conv2D--CubeSphere & 64 & $3\times3$ & (6, 32, 32, 64) & 147,584 \\
    UpSampling2D & & $2\times2$ & (6, 64, 64, 64) & \\
    Concatenate (1) & & & (6, 64, 64, 128) & \\
    Conv2D--CubeSphere & 64 & $3\times3$ & (6, 64, 64, 64) & 147,584 \\
    Conv2D--CubeSphere & 64 & $3\times3$ & (6, 64, 64, 64) & 73,856 \\
    Conv2D--CubeSphere & $vt$ & $1\times1$ & (6, 64, 64, $vt$) & 1,560 \\
    \hline
    \multicolumn{5}{l}{\small $^a$Output shape is (face, $y$, $x$, channels).} \\
    \multicolumn{5}{l}{\small $^b$Number of learned parameters for $t=2$, $v=6$, $c=4$. Total is 2,676,376.} \\
    \end{tabular}
    \end{center}
\end{table}

\section{Designing an ensemble of DLWP models}

The basic ensemble design follows the typical practice used in operational NWP forecasting by including ensemble members with both perturbed initial conditions (ICs) and variations in the model's representation of the atmosphere---the later being  incorporated in NWP ensembles either through the use of several different ``physics" parameterization packages, through a suite of different parameter values with a fixed set of packages, or through the incorporation of stochastic physics.  The perturbed initial conditions and our approach to varying the model representation of the atmosphere are discussed below.

\subsection{Initial condition uncertainty}

The ERA5 dataset includes 10 perturbed ensemble members generated by ensemble data assimilation with 4DVAR \cite{Isaksen2010} to help with uncertainty estimation, and we use these as a convenient set of perturbed ICs for construction of our DLWP ensemble. Unfortunately, this set of ICs is non-optimal, because, unlike the operational ECMWF ensemble \cite{Palmer2018}, singular vectors were not used to select the most rapidly growing initial perturbations.   Moreover, the ERA5 ensemble itself is moderately under-dispersive\footnote{
\tt{https://confluence.ecmwf.int/display/CKB/ERA5\%3A+uncertainty+estimation}}.

\begin{figure}
	\centering
		\includegraphics[width=4in, keepaspectratio]{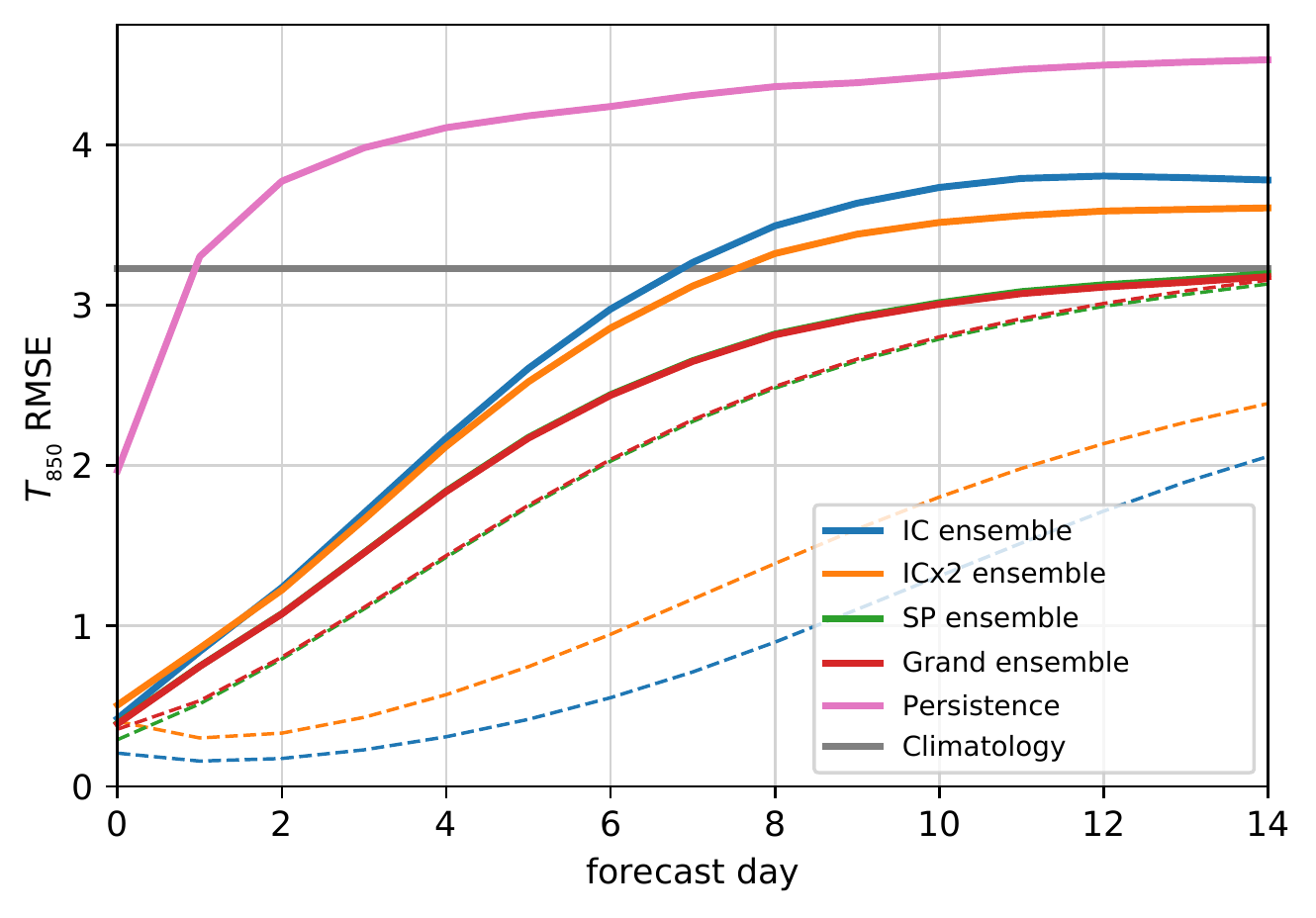}
        \caption{\label{ic} \small RMSE in $T_{850}$ (K) as a function of forecast lead time for DLWP ensembles (solid lines) and corresponding ensemble spread (dashed): IC (blue), IC$\times$2 (orange), stochastically perturbed (SP green) and the grand ensemble (red).  Curves for the SP and grand ensemble are almost identical.  Also shown are the RMSE for persistence (pink) and climatology (gray) benchmarks.}
\end{figure}

Figure~\ref{ic} shows the globally-averaged RMSE of the ensemble mean $T_{850}$ plotted as a function of forecast lead time for four DLWP ensemble strategies.  Also plotted are RMSE reference curves for climatology and persistence. The solid blue curve shows the RMSE of a DLWP forecast generated using the 10 perturbed members of the ERA5 dataset to create a 10-member IC ensemble.
Since the ERA5 IC perturbations do not project strongly on to the most rapidly growing modes, the ensemble spread (computed as the square root of the average over all forecasts of the ensemble variance) actually decreases over the first 36~h of forecast lead time (dashed blue line in Fig.~\ref{ic}). In a conventional NWP model, such a reduction in ensemble spread, which occurs primarily on small spatial scales, can arise from a combination of numerical dissipation and the dispersion of inertia-gravity waves, and an analogous behavior is present in the DLWP model.

Another serious problem with the IC ensemble is that the spread is much smaller than the RMSE of the ensemble mean.  In an ideal ensemble, the joint distribution of the ensemble members would be unchanged if the verifying observations were substituted for any one of the individual ensemble members, and under that assumption the spread and RMSE curves should coincide \cite{Fortin2014}. In an effort to better match the ensemble spread to the RMSE, a 10-member ``IC$\times$2'' ensemble was created, in which the difference in the IC perturbations from the control ERA5 data was doubled.  The spread for the IC$\times$2 ensemble is modestly improved, and over the period 5--14 days the RMSE is modestly reduced (orange curves) relative to the original IC ensemble. Note that the reduction in initial ensemble spread over the first 36~h is more rapid in the IC$\times$2 ensemble, providing further evidence that the variations in these initial condition do not strongly project on the structure of the most rapidly growing perturbations.

\subsection{Uncertainty in the representation of the atmosphere}

The learnable weights for convolutions in the CNN are initialized as small random values, and we can exploit this randomness by repeatedly retraining with different initial seeds to produce a family of DLWP models with slightly different final weights. The models in this family are capable of making approximately equally-skillful forecasts, but with enough statistical independence to produce a good ensemble.  \citeA{Scher2020} pursued a similar strategy, training and re-training models to both emulate a simple GCM and to forecast the real atmosphere with data from ERA5. 

As part of its holistic estimation of the next atmospheric state, our CNN based DLWP model effectively captures physical processes that are parameterized in operational NWP models.  For example, as noted in WDC20 (see also Fig.~\ref{ensemble}c,d), our DLWP model does an excellent job of forecasting 2-m temperatures, including their diurnal cycle, without using any explicit parameterization of boundary layer processes, and without including most of the meteorological fields that would be used in such parameterizations.  We therefore refer to the ensembles in which the CNN filter coefficients are randomly perturbed during training as  ``stochastically perturbed'' (SP) ensembles, although the nature of the induced stochastic variations differs from that in operational NWP.

\begin{figure}
	\centering
		\includegraphics[width=4in, keepaspectratio]{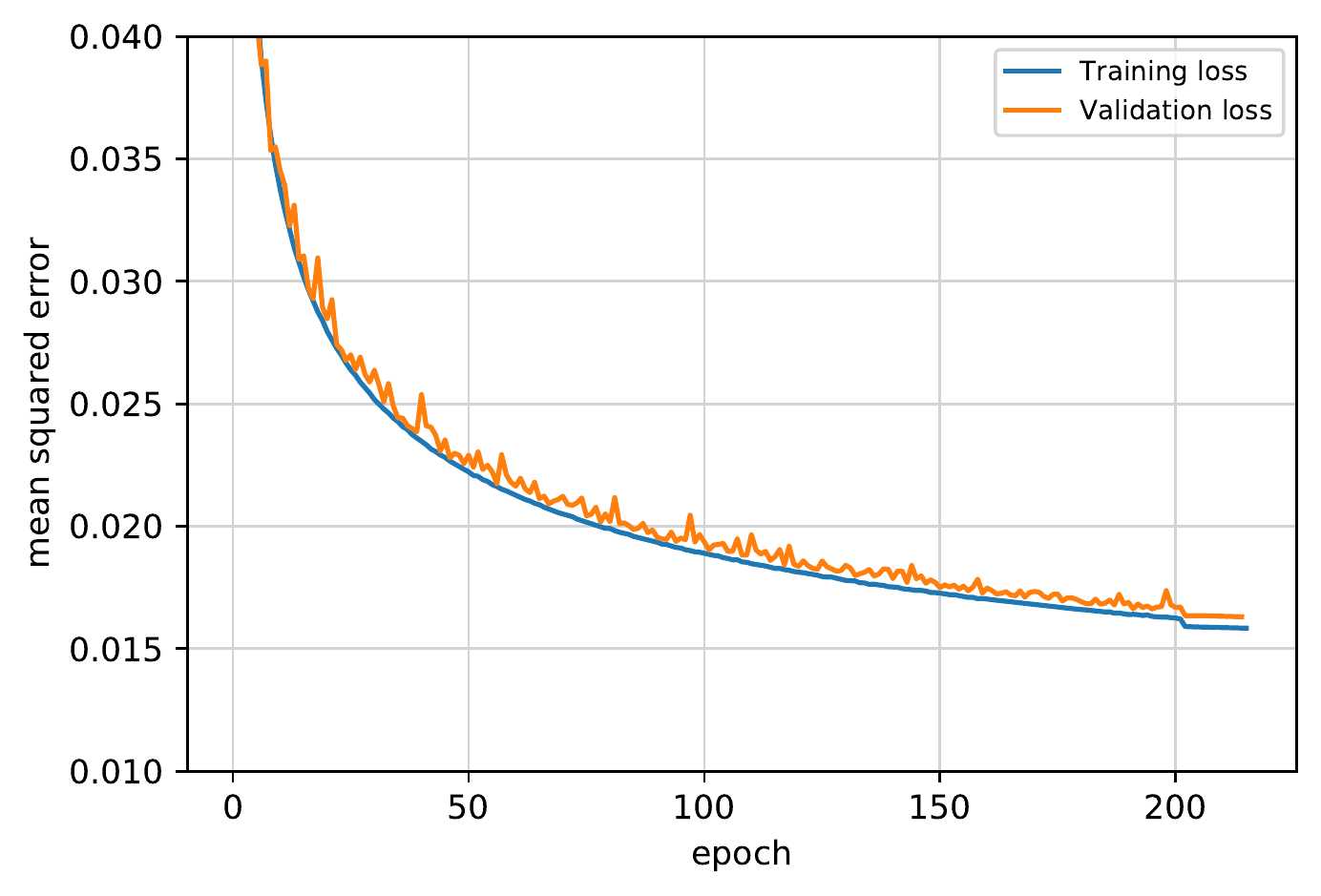}
        \caption{\label{learning} \small Example of the CNN learning curve for a representative DLWP model. The loss, which is the mean-squared-error over all target outputs of the model, evaluated on the training (validation) set is shown as a function of training epoch number in blue (orange). The optimizer was switched from Adam to a standard SGD optimizer after 200 epochs, producing an abrupt small decrease in the loss function on both data sets and more uniform values on the validation set.}
\end{figure}

Rather than completely retrain each member of our SP ensemble from new random seeds, we gained efficiency by using intermediate results produced during the training process.  Our DLWP model is trained using the adaptive learning scheme Adam \cite{Kingma2014}. 
Figure~\ref{learning} shows the learning curve for our loss function (mean squared error, see WDC20) as a function of the training epoch number for a training cycle representative of that for one of our DLWP ensemble members.  As expected, the error on the training set decreases smoothly, while the error on the validation set oscillates much more, strongly suggesting the learned weights in the model undergo nontrivial changes over  each training epoch. The variations induced by these changes in the weights turn out to be sufficient to provide many useful members in a SP ensemble, and we exploited these variations as follows.

After at least 100 training epochs, we selected multiple potential ensemble members from a single training cycle by checkpointing every 10 epochs and saving the model's weights at each checkpoint.  Using these checkpointed weights, we tested each resulting model's utility for S2S forecasts by evaluating its average global T$_{850}$ anomaly correlation coefficient (ACC) score at 4-week lead time from twice-a-week forecasts over the full four-year validation set.  The 4-week T$_{850}$ ACC scores for the checkpointed models were often equally good (or occasionally even better) than that for final trained model\footnote{Although the final trained model had the lowest validation set loss during training, the training loss function is based solely on the RMSE over the first 24 hours of forecast lead time; it does not depend on the performance of a 4-week forecast.}, and there was nearly as much ensemble spread between the various checkpoints of one training cycle as there was between models generated by separate training cycles.

Our SP ensemble used a total of 32 slightly different DLWP models generated by eight training cycles with four members drawn from each cycle.  The members selected from each cycle had the best 4-week T$_{850}$ ACC scores, although three of the 32 members selected in this fashion required further refinement because in a few individual forecasts they produced fields with unrealistic structures even though their numerical values remained bounded within reasonable limits.  Those three members were further trainined for two more epochs using a re-initialized stochastic gradient descent (SGD) optimizer, which consistently fixed the issue with nonphysical solutions.  The SGD cycles also lowered the training and validation loss slightly, as exemplified over the last 15 epochs of the learning curve in Fig.~\ref{learning}.  

The RMSE and spread of the SP ensemble is plotted in Fig.~\ref{ic}.  It clearly outperforms the IC and IC$\times 2$ ensembles, having lower RMSE and a much better RMSE-spread relationship.  The spread is roughly 80\% of RMSE at a forecast lead time of 8 d and approaches 95\% by 14 d.  At 14 d, the RMSE of the SP ensemble remains slightly better than climatology, whereas the RMSE for the IC and IC$\times 2$ ensembles begins to exceed climatology between 7 and 8 d. \citeA{Scher2020} compared the performance of initial-condition and retrained (i.e. SP) neural network ensembles.  For the 500-hPa forecasts using the less accurate DLWP model of \citeA{Weyn2019}, they found a relationship between the RMSE and spread in their SP ensemble roughly similar to that in Fig.~\ref{ic}, although in their case the spread matched the RMSE at the end of their 5-day forecast and the RMSE 500-hPa was roughly twice as large as that for our current model at the same 5-day forecast lead time (compare Fig.~\ref{ensemble}a with their Fig.~4a). In contrast to our IC ensembles, theirs was generated using singular vectors and produced only slightly less spread than their SP ensemble.  Finally, as in our results, the RMSE of their IC ensemble mean was higher than the RMSE for their SP ensemble mean.

\citeA{Scher2020} did not report results for a grand ensemble consisting of both IC and SP perturbations, but given the superior ensemble spread they obtained using singular vectors, such a grand ensemble might have performed substantially better than either of the individual IC or SP ensembles.  In our case, a 320-member grand ensemble constructed by applying the suite of 32 DLWP models with slightly different weights to each of the 10 IC perturbations performs only very slightly better than the SP ensemble alone at 14-d forecast lead times: the RMSE and spread curves (red) for the grand ensemble almost perfectly overlap those for the SP ensemble (green) in Fig.~\ref{ic}.  Nevertheless,
after bias correction (see section \ref{section-bias}) the 320-member grand ensemble does perform better than the SP ensemble at longer S2S forecast lead times.

\subsection{The control member}

When considering the effectiveness of ensemble forecasts, it is useful to compare the ensemble mean to a single control member. For example, in comparison to the other ensemble members, the ECMWF control forecast is run at higher horizontal (9~km) and vertical (137 levels) resolution, and without perturbations to the initial conditions. This control forecast might nominally be expected to perform better than a typical ensemble member that must be run at lower resolution because of computational constraints.
Our control forecast is trained to better minimize the loss function using a Adam optimizer and learning rates that decrease as the weights approach their optimum values.  In particular, the learning rate starts at $10^{-3}$, but once the validation-set loss does not decrease for 20 epochs, the learning rate of the optimizer is reduced by a factor of 5. This continues (up to a minimum learning rate of $10^{-6}$) until a criterion of no reduction in validation loss after 50 epochs is met. The result is a model with weights that better minimize the loss function and produces good forecasts, although in contrast to the ECMWF high-resolution control, the model used for our control forecast does not have significant advantages relative to the other ensemble members. The control forecast is initialized with the control ERA5 reanalysis data.  

\subsection{\label{section-bias} Correcting model bias}

Unlike global climate models, NWP models are designed to make accurate predictions over relatively short forecast lead times without worrying about certain physical constraints, such as global radiative balance, that would be necessary for long-term climate simulations. 
As a consequence, NWP models are typically subject to systematic drift in long-term (including sub-seasonal) forecasts. For example, previous versions of the ECMWF S2S ensemble have been shown to develop pronounced spatially-dependent patterns of mean model drift on time scales of 1--4 weeks \cite{Vitart2004,Weigel2008}.  
To compensate for this model bias, all of the major weather prediction centers produce reforecasts, or hindcasts, using their S2S ensemble prediction systems \cite{Vitart2017}, and use these reforecasts to calibrate the operational forecast products. 
As an example, \citeA{Vitart2004} computed the bias for a given calendar date by performing hindcasts using the full ensemble initialized on the same calendar date in each of 12 previous years, and removed this bias from the forecasts in a post-processing step.

\begin{figure}
	\centering
		\includegraphics[width=\textwidth, keepaspectratio]{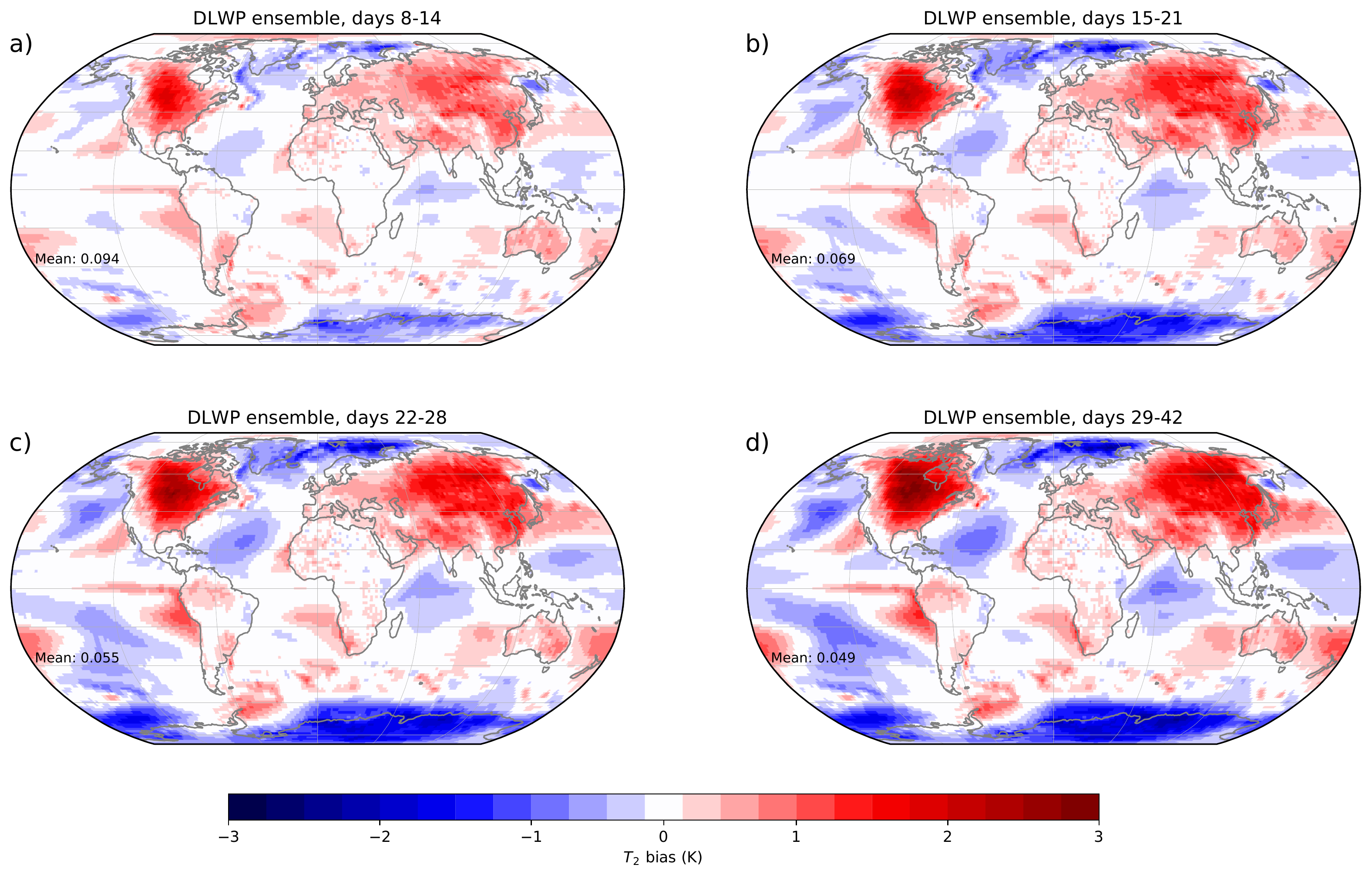}
        \caption{\label{bias-map} \small Bias in 2-m temperature (ensemble-mean $-$ observation) averaged over 25 years of re-forecasts from all members of the DLWP SP ensemble: one-week averages for forecast lead times of a) 2 weeks, b) 3 weeks, c) 4 weeks, and d) a 2-week average for the week 5--6 forecast.}
\end{figure}

Using a strategy similar to that in \citeA{Vitart2004}, we bias corrected each field by first computing ensemble reforecasts twice weekly for each DLWP SP ensemble member (and the control) on the same set of calendar dates spanning the years 1991--2015, which includes the training set and most of the validation set, but does not bleed into the 2017-2018 test set.  Then, for all reforecast dates, each member's spatially varying bias is calculated as the average of its bias on that date over the 25~year period.  The bias for a specific forecast date in the test set is taken as the average reforecast bias over all available calendar dates spanning the 28-day interval centered on that forecast date, and this bias is subtracted from the forecast produced by each of the corresponding ensemble members.

Figure~\ref{bias-map} shows spatial patterns of the annual-average model bias in 2-m temperature for the 32-member DLWP SP ensemble at forecast lead times up to 6~weeks. Warm biases are present over the northern hemisphere land masses, along with a cold bias over Antarctica. There are also warm biases in subtropical regions commonly dominated by marine stratocumulus clouds off the Pacific coasts of North and South America. These biases gradually amplify as the forecast lead time increases, although the globally averaged spatial-mean bias (noted in each panel) decreases at longer lead times.  The tendency of increasing local biases to better cancel in the global mean at longer lead times is interesting and perhaps surprising because the model is only trained to minimize $T_2$ errors over the first 24 hours of the forecast---no global energy-balance constraints are imposed.

Bias correction has a positive impact on the control forecast and on the IC, SP, and the grand ensembles.  Although the RSME and spread of the SP and grand ensembles are almost identical over the first 14 days, at longer lead times, and particularly after bias-correction, the grand ensemble is clearly superior to the SP ensemble (not shown). The performance of the grand ensemble will, therefore, be our focus throughout the remainder of this paper.  

In addition to the persistence and climatology benchmarks, which serve as a baselines that must be exceeded by any skillful forecast, we will also compare our results against the state-of-the-art ECMWF 50 member S2S ensemble and a higher resolution ECMWF control simulation \cite{Vitart2017}.  Errors are computed with respect to ERA5 data that is downloaded at 1 degree resolution, transformed onto our cube-sphere grid, and then transformed back to a 1.5$\times$1.5 latitude-longitude grid.  Our DLWP forecasts are transformed to the same 1.5$\times$1.5 degree grid for the computation of all forecast metrics.  The archived ECMWF S2S forecasts, available on a 1.5$\times$1.5 degree grid, are first transformed to the cube sphere and then back to the
1.5$\times$1.5 degree analysis grid because this procedure removed discrepancies in model terrain thereby improving the ECMWF error metrics for $T_2$.
Bias correction was also performed on the ECMWF S2S control and ensemble forecasts on the 1.5$\times$1.5 degree grid, with the methodology following that of the operational ECMWF forecasts. 
This correction is very similar to the bias correction applied to our DLWP model, but with a few differences: the last 20 years of re-forecasts are used instead of a fixed period of 25 years; ten ensemble members with perturbed IC and physics are run for each re-forecast; and only the forecasts for dates within one week, instead of 28 days, of the target operational forecast issue date are used.

\subsection{Summary}

The following summarizes the construction of the DLWP grand ensemble.
\begin{enumerate}
    \item Eight distinct training cycles of the DLWP CNN were produced with different random seeds as a first step in generating 32 stochastically perturbed (SP) models.
    \item Four checkpoints during each of the eight training cycles were selected based on $T_{850}$ ACC skill as individual SP ensemble models.
    \item The model associated with each checkpoint was run on the validation data set to produce 416 four-week forecasts, which were then manually inspected for forecast quality. Any model that displayed irregularities was further trained with an SGD optimizer. The collection of models given by these checkpoints formed the 32-member SP ensemble of models. (The SGD optimizer was used in 3 of the 32 SP ensemble members.)
    \item Each of the 32 SP models was run with each of the 10 initial conditions (ICs) given by the perturbed reanalyses in the ERA5 product to yield the 320-member grand ensemble.
    \item A single control DLWP model was trained slightly differently, by periodically reducing the Adam optimizer learning rate. 
    \item The mean model bias for re-forecasts in the period 1991--2015 was computed for the control and each SP model. That bias was removed from all the 2017-2018 test-set forecasts.
\end{enumerate}

\begin{table}[t]
    \caption{\label{ensemble-table}  \small  Comparison of key attributes of our DLWP ensemble and those of the state-of-the-art ECMWF ensemble for extended-range forecasting.}
    \begin{center}
    \begin{tabular}{ |c|c|c| } 
        \savehline
        \rowcolor{gray}
         & DLWP & ECMWF \\ 
        \savehline
        Atmospheric fields & \cellcolor{CellRed} 6 2-D variables & \cellcolor{CellGreen} 9 prognostic 3-D variables; 91 vertical levels \\ 
        \savehline
        Horizontal resolution & \cellcolor{CellRed} ~150 km &\cellcolor{CellGreen}  ~18 km (~36 km after day 15) \\ 
        \savehline
        Atmospheric physics & \cellcolor{CellRed}3 prescribed inputs & \cellcolor{CellGreen} Many physical parameterizations \\
        \savehline
        Coupled models & \cellcolor{CellRed}None & \cellcolor{CellGreen} Ocean, wave, and sea ice models \\
        \savehline
        Initial condition perturbations & \cellcolor{CellRed}10 (ERA5 uncertainty) & \cellcolor{CellGreen} 50 (SVD/4DVAR) \\
        \savehline
        Model perturbations & ``Stochastic'' CNN weights & Stochastic physics \\
        \savehline
        Ensemble members & \cellcolor{CellGreen} 320 (+control) & \cellcolor{CellRed}50 (+control) \\
        \savehline
    \end{tabular}
    \end{center}
\end{table}

Finally, a tabular comparison between our DLWP ensemble and the current state-of-the-art ECMWF ensemble is provided in Table~\ref{ensemble-table}. In most regards, the ECMWF ensemble is superior, with higher resolution, coupled ocean and wave models, complete physics, more atmospheric variables, and better initial condition perturbations. However, our DLWP ensemble consists of far more ensemble members, with 320 plus control, compared to only 50 plus control for the ECMWF ensemble.

\section{Skill of the DLWP at short lead times} 

Before discussing the performance of the DLWP ensemble on S2S time scales we briefly assess the qualitative skill of the model in a short deterministic forecast.  We then assess the quantitative skill of the DLWP control and grand ensemble in two-week forecasts relative to the state-of-the-art 50-member ECMWF S2S ensemble and the standard benchmarks of climatology and persistence. 

\begin{figure}
	\centering
		\includegraphics[width=\textwidth, keepaspectratio]{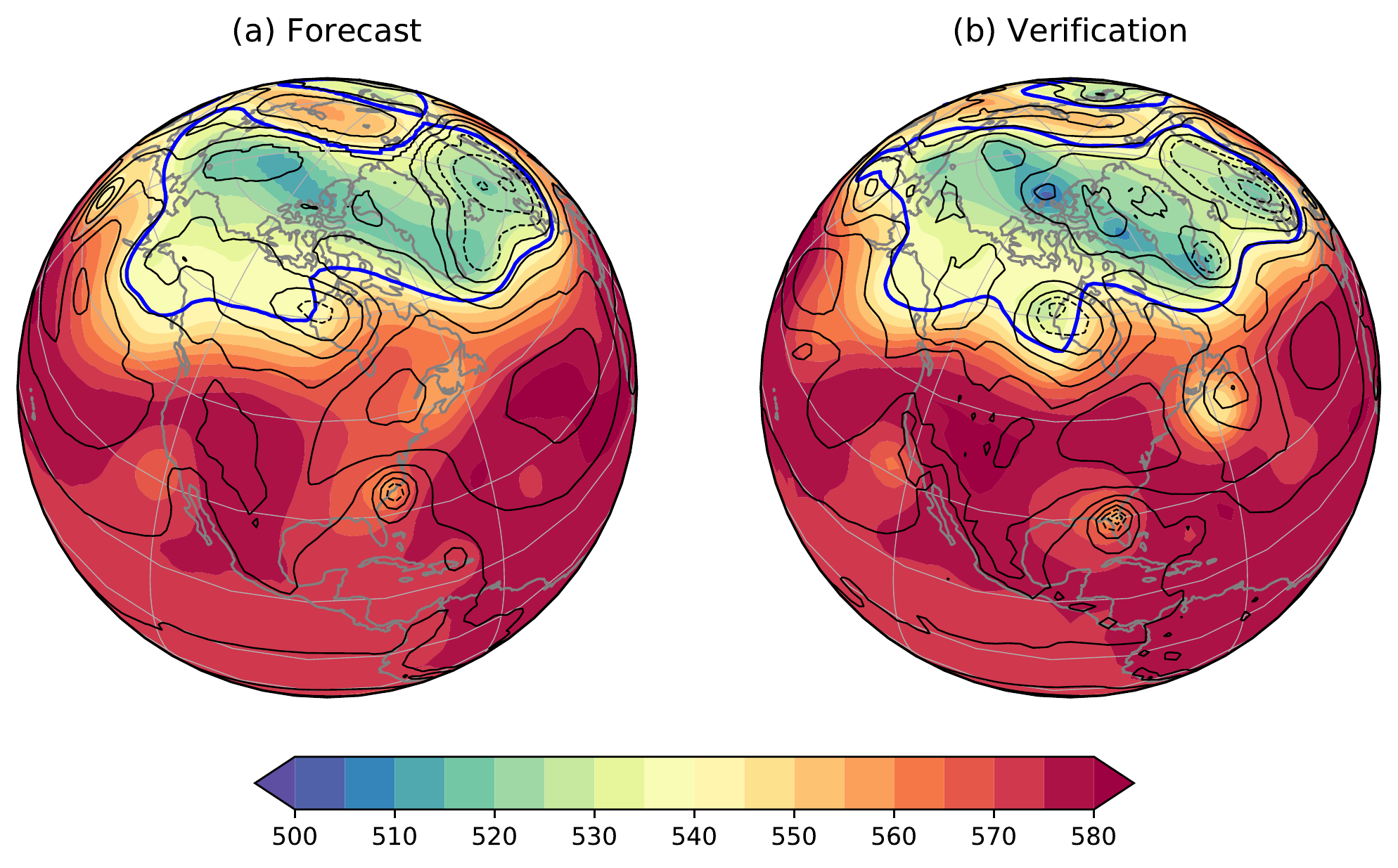}
        \caption{\label{irma} Fields of $Z_{500}$ (color contours) and $Z_{1000}$ (black contours at 100 m intervals with negative values dashed) for (a) a 4.5-day forecast and (b) the verification on 12:00 UTC, 11 September 2017.  The blue curve is the 540-dm contour for $Z_{500}$.}
\end{figure}

Figure~\ref{irma} compares a 4.5-day global forecast of $Z_{1000}$ and $Z_{500}$ with the verifying analysis for 12:00 UTC, 11 September 2017, when  hurricane Irma was located in the southeastern US. Irma is farther north and weaker in the DLWP forecast, but still reasonably well represented for a 4.5-day forecast of such a small-scale disturbance.  Other features, such as the 500-hPa cutoff low west of California and the pair of short waves in the Gulf of Alaska and west of Hudson Bay are also reasonably represented.  On the other hand, the  500-hPa cutoff low over Novia Scotia is much weaker and farther west than in the verification, and the DLWP forecast does not develop the associated surface cyclone. The closed surface low in the DLWP forecast over the Dominican Republic is the model's forecast for hurricane Jose, which is stronger and farther south than in the verification.  While neither perfect, nor the equal of a state-of-the-art NWP operational forecast, on the balance the DLWP forecast is arguably still impressive given that it is computed at 1.4$^{\circ}$ resolution using just 6 prognostic variables, each defined on a single spherical shell. 

Turning to the quantitative verification of the first two weeks of forecast lead time, our cases are chosen to match the available forecast initialization times from the operational S2S forecast runs at ECMWF.  The DLWP model is therefore tested on 208 forecasts initialized twice weekly starting at 00 UTC 2 January 2017, followed by the 5th, 9th and 12th of January, and so on, through the end of 2018. 

\begin{figure}
	\centering
		\includegraphics[width=\textwidth, keepaspectratio]{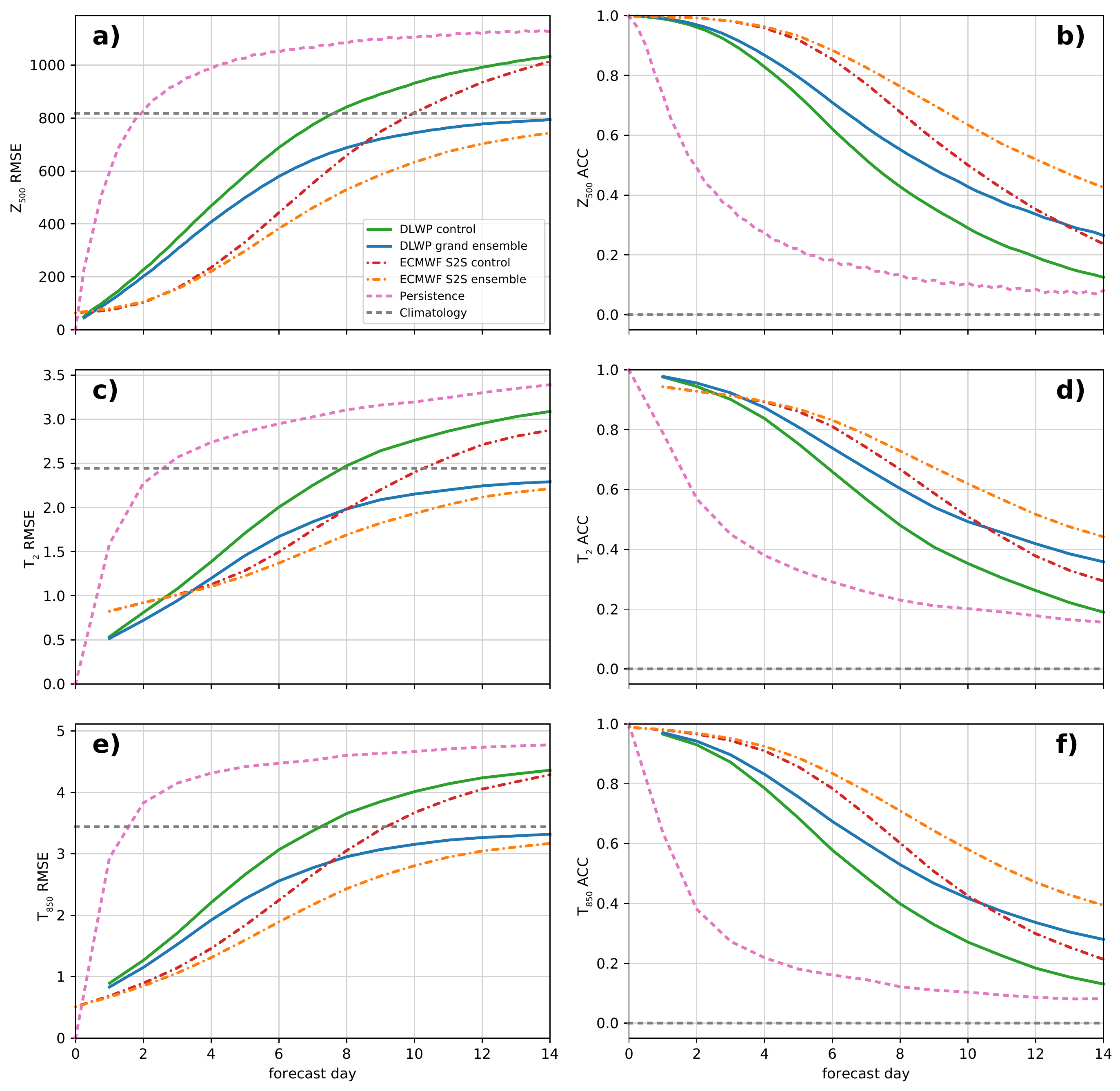}
        \caption{\label{ensemble} \small Forecast error as a function of time for the DLWP control member (green) and grand ensemble mean (blue), the ECMWF S2S control (dot-dashed red) and ensemble mean (dot-dashed orange), along with persistence (pink) and climatology (gray) benchmarks for twice-weekly forecasts during 2017--2018. Panels are $Z_{500}$: (a) RMSE (m) and (b) ACC;
        daily-averaged $T_2$: (c) RMSE (K) and (d) ACC; $T_{850}$ at 00 UTC: (e) RMSE (K) and (f) ACC.
        The error is area-weighted in latitude and globally-averaged.}
\end{figure}

RMSE scores for $Z_{500}$ are compared in Fig.~\ref{ensemble}a. The DLWP grand ensemble remains superior to climatology through 14 days, though unsurprisingly, its error exceeds that of the ECMWF S2S ensemble.  Both the DLWP and ECMWF control forecasts perform worse than their respective ensembles, and the control-to-ensemble improvement is qualitatively similar for both systems.  At lead times beyond 9 days, the DWLP ensemble performs better than the ECMWF control.  Similarly qualitative behaviors are apparent for the $Z_{500}$ anomaly correlation coefficient (ACC) scores shown in Fig.~\ref{ensemble}b, although the ACC score for the ECMWF remains superior to that of the DLWP ensemble until almost day 13.  The persistence forecast is grossly inferior.

Turning now to the daily-averaged\footnote{Only the daily-averaged $T_2$ field was available on the ECMWF archive.  As shown in WDC20, the DLWP model does capture diurnal temperature variations.} surface temperature field, which has a far more complex structure than $Z_{500}$, Fig.~\ref{ensemble}c,d show the performance of 
the DLWP models relative to the ECMWF system remains similar to that for $Z_{500}$. The  ECMWF S2S ensemble gives the best results, with its RMSE,  together with that of the DLWP grand ensemble, remaining below the climatological benchmark through 14 days.   Note that the ECMWF product uses slightly different initial conditions, which explains the substantial error in the ECMWF RMSE at early lead times. The ACC of the DLWP ensemble again becomes superior to the ECMWF control at long lead times (after 11 days). 
The ability of the model to correctly forecast surface temperatures using the same set of 3$\times$3 filters over both land and ocean highlights the capability of the deep learning approach to capture processes that require complex physical parameterizations in conventional NWP models with no more information than the land-sea mask and the terrain elevation.

Error metrics for a third field, $T_{850}$, are plotted in Fig.~\ref{ensemble}e,f. This is a more difficult field to forecast in the sense that the RMSE for persistence drops below the climatological skill threshold at shorter lead times than for $Z_{500}$ or $T_2$, and the ACC score for persistence drops below 0.4 in just two days. Nevertheless, the RMSE for the DLWP and ECMWF ensembles again remains below climatology for 14 days, with the ECMWF ensemble performing the best.  The errors in the DLWP ensemble also drop below those in the ECMWF  control at earlier lead times than for the other fields: 7 days for RMSE and 10 days for ACC.

\section{Extending the forecasts to the S2S range}

\subsection{\label{s2sacc}Ensemble-mean anomaly correlations}

\begin{figure}
	\centering
		\includegraphics[width=\textwidth, keepaspectratio]{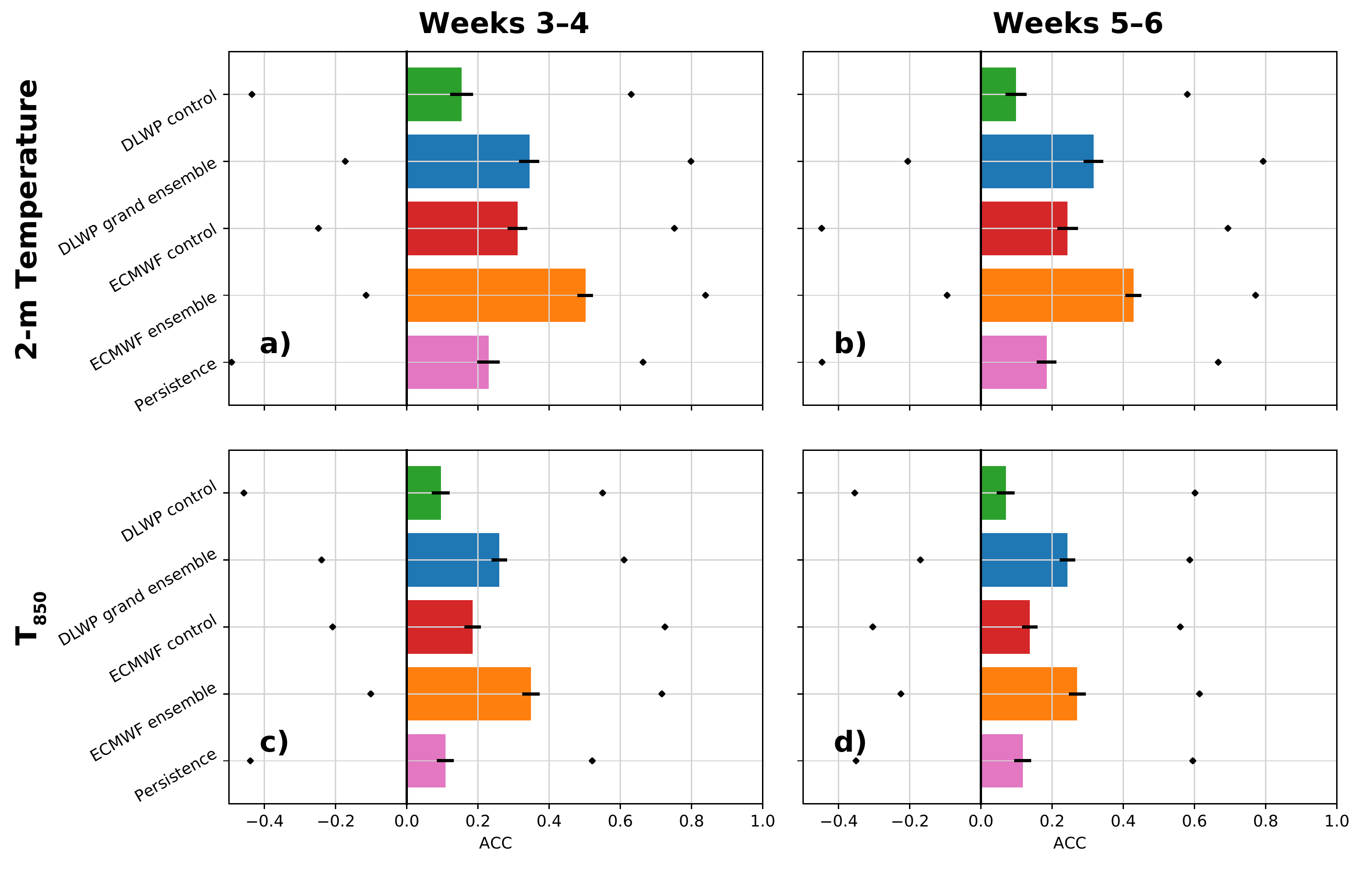}
        \caption{\label{s2s-acc-var} \small Anomaly correlation coefficient of 2-week-averaged forecasts from the DLWP control (green) and grand ensemble mean (blue), the ECMWF S2S control (red) and ensemble mean (orange), and persistence (pink) for forecasts made twice weekly in 2017--2018. Panels show $T_2$ for (a) weeks 3--4, (b) weeks 5--6, and $T_{850}$ for (c) weeks 3--4, (d) weeks 5--6. Scores are area-weighted in latitude and globally-averaged.  Black lines on each bar represent the 95\% confidence interval computed using bootstrapping with 10,000 iterations. The black dots show the lowest and highest scores among the 208 forecasts.}
\end{figure}

Globally and temporally averaged ACC scores for $T_2$ and $T_{850}$ at forecast lead times of 3--4 and 5--6 weeks are compared in Fig.~\ref{s2s-acc-var} for the DLWP and ECMWF models, along with the persistence benchmark, for the 2017-2018 test set. 
Persistence forecasts are computed by averaging the observed anomalies from the 14 days prior to the forecast initialization date.  At both lead times, scores are higher for $T_2$ than for $T_{850}$, reflecting greater memory in the system for near-surface temperatures than those aloft.  One source of this memory are 
sea surface temperatures, which are closely tied to $T_2$ over the oceans.  Unlike the ECWMF S2S model, our DLWP model does not currently include coupling with the ocean, and as a consequence, our $T_2$ forecast over the ocean is essentially a proxy forecast for SST. This may be one reason why the DLWP control is the only model in Fig.~\ref{s2s-acc-var}a,b that performs worse than persistence.  Nevertheless for $T_2$ forecasts at both lead times, the DLWP ensemble is superior to both persistence and the ECMWF control.  Unsurprisingly, the ECMWF ensemble gives the best results at both lead times, with an averaged ACC of roughly 0.5 for $T_2$ at 3--4 weeks.
Turning to $T_{850}$, the performance of the DLWP ensemble relative to the other  forecasts is better than those for $T_2$.  The DLWP ensemble is again superior to the ECMWF control at both lead times, and more impressively, at weeks 5--6 it is in a statistical tie with the full ECMWF S2S ensemble in the sense that the 95\% confidence intervals for the forecasts overlap (black bars in Fig.~\ref{s2s-acc-var}d).

ACC scores for the best and worst individual forecasts for each period are shown by the black dots in Fig.~\ref{s2s-acc-var}.  The ECMWF ensemble has the ``best" worst forecasts, or the least susceptibility to bust forecasts, except for 5--6-week $T_{850}$, for which its ACC of -0.22 is worse than the $-0.18$ value for DLWP ensemble.  At weeks 5--6, the best individual forecasts for both the DLWP and ECMWF ensembles have ACC scores of about 0.8 for $T_2$ and about 0.6 for $T_{850}$. The globally averaged ACC for the 3--4-week $T_2$ forecasts exceeds 0.5 roughly 25\% of the time for the DLWP grand ensemble and 50\% of the time for the ECMWF ensemble.


\begin{figure}
	\centering
		\includegraphics[width=\textwidth, keepaspectratio]{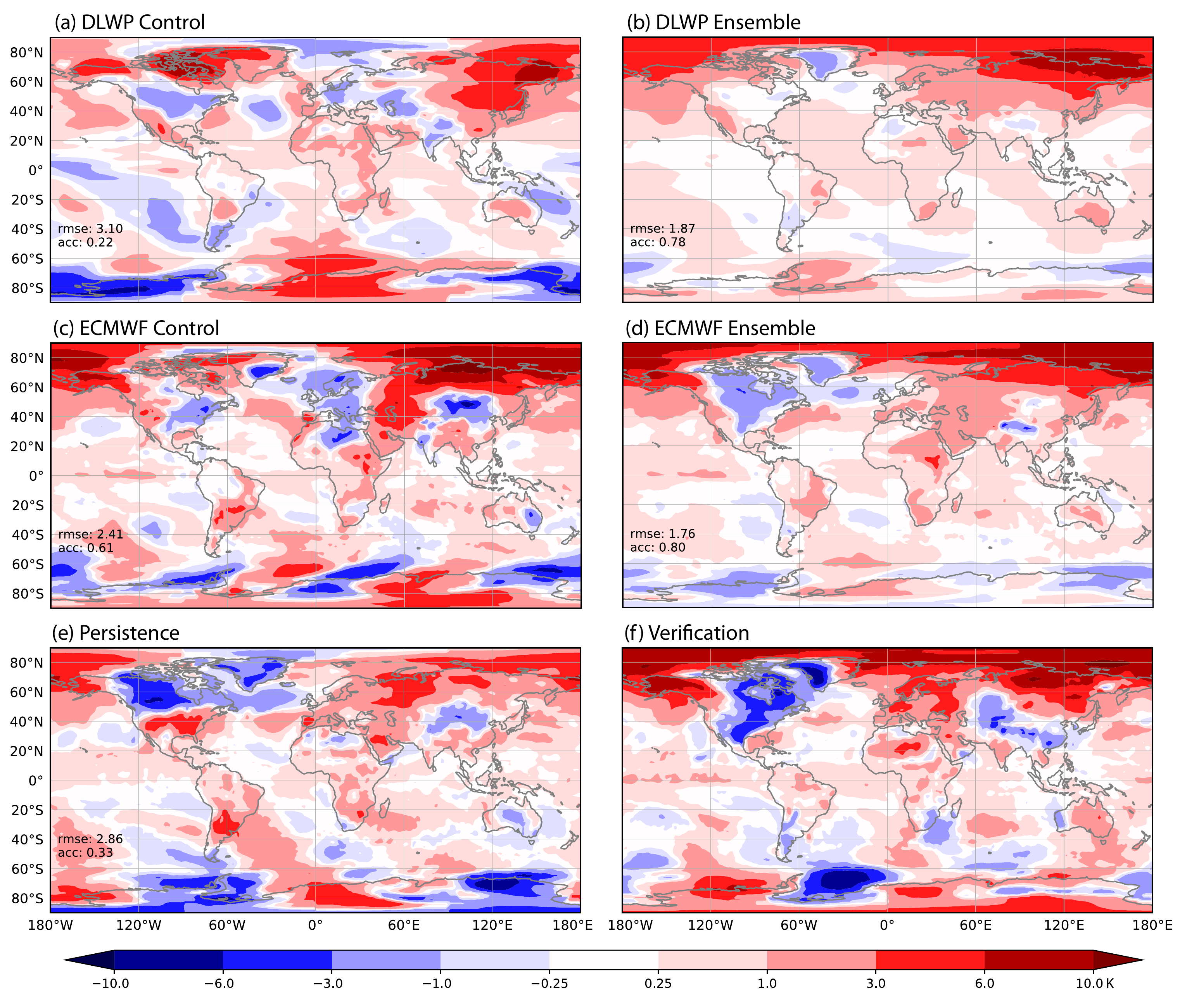}
        \caption{\label{anomaly-sept-3-4} \small Predicted 2-week average anomalies in $T_2$ relative to climatology for the period 3--4 weeks after 27 September 2018. Forecasts are from DLWP (a) control and (b) grand ensemble, ECMWF S2S (c) control and (d) ensemble, and (e) persistence; (f) is the verification.  The `rmse' and `acc' numbers are global averages; the rmse is the root-mean-squared error of the \emph{anomalies}.}
\end{figure}

 An example of a good (though not the best) 3--4-week $T_2$  anomaly forecast for both ensembles is shown in Fig.~\ref{anomaly-sept-3-4}.  These forecasts were initialized on 27 September 2018.  The intensity and spatial variability in the DLWP control forecast are grossly similar, although modestly weaker and smoother, than those in the ECMWF control, demonstrating that the DLWP forecast is not simply approaching a smooth climatology at long forecast lead times.  Similarly, the forecast from the 320-member DLWP grand ensemble exhibits much of the intensity and spatial variability in the 50-member ECMWF ensemble forecast.\footnote{Although the horizontal resolution of the ECMWF S2S ensemble members is 31 km beyond lead times of 15 days, the data are archived at coarser resolution and all forecasts are displayed on a 1.5$^{\circ}$ $\times$ 1.5$^{\circ}$ latitude-longitude grid.}

Anomalies that both verify (Fig.~\ref{anomaly-sept-3-4}f) and are common to the DLWP (Fig.~\ref{anomaly-sept-3-4}a) and ECMWF S2S (Fig.~\ref{anomaly-sept-3-4}b) ensembles include cold in Greenland and warmth in eastern Australia, eastern Siberia and over the adjacent Arctic Ocean.  One place where the ECMWF S2S ensemble clearly out performs the DLWP ensemble is over North America, where it better captures the observed large and intense cold anomaly.  Another important superiority of the ECMWF ensemble lies in its forecast of a developing El Ni\~no over the equatorial Pacific, although the absence of the  El Ni\~no in the DLWP ensemble forecast is not particularly surprising because it does not include any oceanic data.

\subsection{\label{probscore}Ensemble probabilistic scores}

Having just examined the ACC scores of the ensemble mean, we now investigate the performance of ensemble-produced probabilistic forecasts, specifically the continuous ranked probability score (CRPS) and the ranked probability skill score (RPSS).

\subsubsection{Continuous ranked probability score}

Denoting the ensemble probability distribution function of a forecast for some variable $x$ as $\rho(x)$, the cumulative distribution function (CDF) associated with $\rho$ is 
\begin{equation}
	P(x) = \int_{-\infty}^{x} \rho(y) \, dy.
\end{equation}
If the verifying value occurs at $x_a$, a CDF for the observation may be defined as
\begin{equation}
	P_a(x) = H(x - x_a),
\end{equation}
where
\begin{equation}
	H(s) = 
	\begin{cases}
		0 & s < 0 \\
		1 & s \geq 0
	\end{cases}
\end{equation}
is the Heaviside function. The CRPS may then be defined as \cite{Hersbach2000}:
\begin{equation}
	\text{CRPS} = \text{CRPS}(P, x_a) = \int_{-\infty}^{\infty} \left[P(x) - P_a(x)\right]^2 dx.
\end{equation}
The CRPS score penalizes both overly narrow (confident) forecast distributions that verify incorrectly and overly broad (uncertain) forecast distributions, regardless of the accuracy of the ensemble mean. 
The CRPS has several desirable properties. 
First, it is a proper statistical score \cite{Gneiting2007}, meaning that the CRPS is optimized for a forecast which predicts the correct probability distribution of a predicted variable. 
Second, in contrast to categorical scores, such as the RPSS (see Section~\ref{RPSS}), it accounts for information across all possible values of $x$. 
Finally, the CRPS reduces to the mean absolute error for a single deterministic forecast, allowing the performance of ensemble and deterministic forecasts to be easily compared.  

\begin{figure}
	\centering
		\includegraphics[width=0.7\textwidth, keepaspectratio]{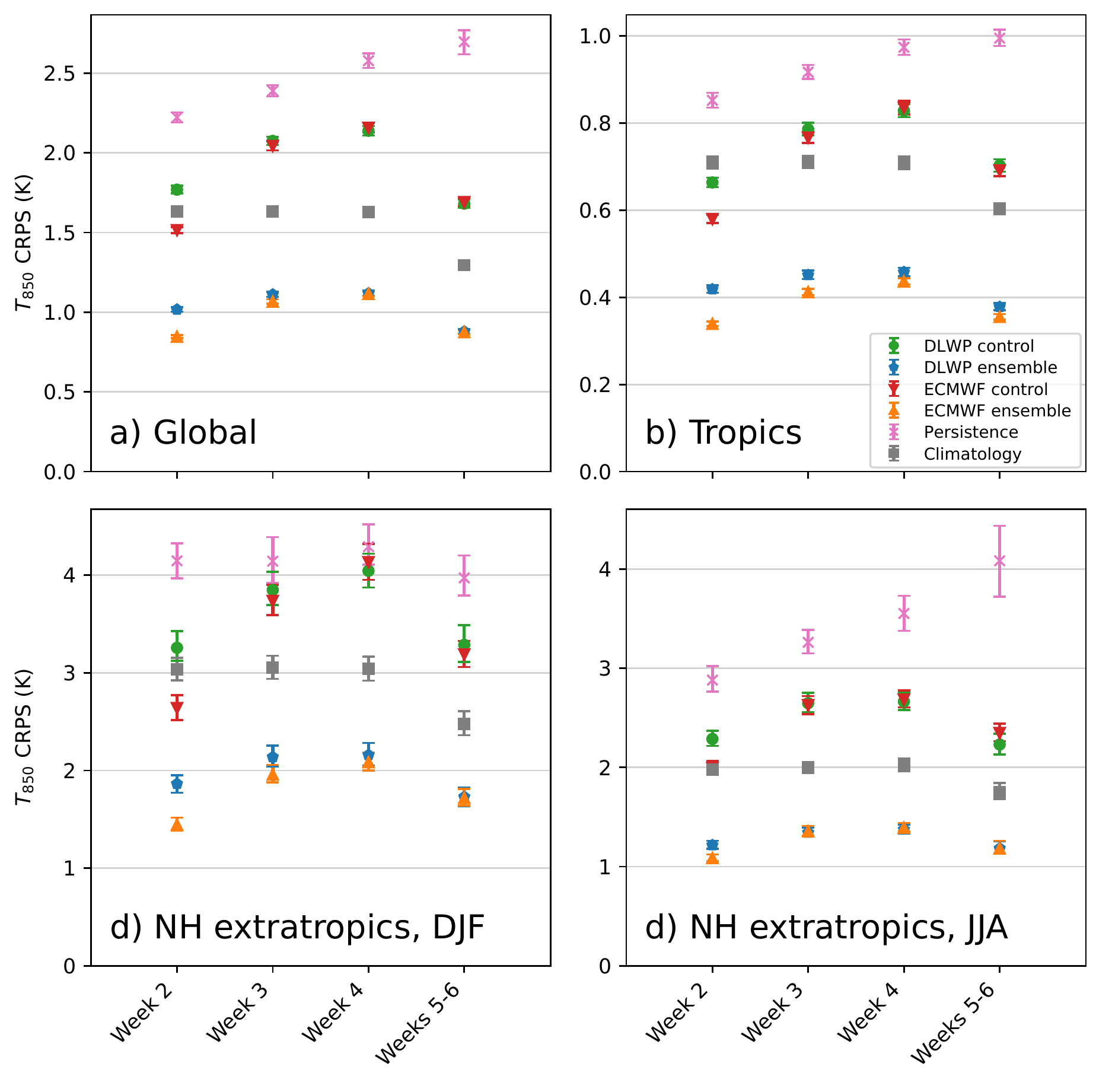}
        \caption{\label{crps} Continuous ranked probability score (CRPS; lower is better) in $T_{850}$ from the DLWP control (green circles), DLWP grand ensemble (blue pentagons), ECMWF S2S control (red downward pointing triangles), ECMWF ensemble (orange upward pointing triangles), persistence (pink crosses), and climatology (gray squares), as a function of averaged forecast-lead time. Panels show: a) Global average, annual mean; b) average over the tropics (20$^{\circ}$S -- 20$^{\circ}$N), annual mean; c) average over the northern hemisphere extra-tropics (30$^{\circ}$N -- 90$^{\circ}$N), mean of forecasts initialized in DJF; d) average over the NH extra-tropics, JJA. Error bars correspond to the 95\% confidence interval determined by bootstrapping with 10,000 samples. }
\end{figure}

Figure~\ref{crps} compares CRPS in $T_{850}$ for forecasts from the DLWP and ECMWF S2S ensembles and control forecasts, along with persistence and climatology benchmarks, averaged over one week for lead times of 2, 3 and 4 weeks, along with a two-week average for a lead time of 5--6 weeks.  As with our earlier results for the $T_{850}$ RMSE and ACC of the ensemble mean (Fig.~\ref{ensemble}),  the ECMWF ensemble clearly gives better week-2 CRPS scores than the DLWP ensemble.  
At week 3, when deterministic forecast skill from initial conditions has largely eroded, the ECMWF ensemble continues to give a slightly better global mean result. But by week 4 and weeks 5--6, the DLWP ensemble has caught up and is essentially tied with ECMWF 
 (Fig.~\ref{crps}a).  At lead times of three weeks or longer, the next best global-mean forecasts are given by climatology, which out-performs the ECMWF and DLWP control forecasts, which are in turn better than persistence.  Note that all CRPS scores except persistence improve significantly from week 4 to weeks 5--6 due to the longer two-week averaging window.

Focusing on the tropics, from 20$^{\circ}$S to 20$^{\circ}$N, the CRPS for all models are substantially improved, with numerical values roughly 0.4 times the corresponding global results (Fig.~\ref{crps}b).  The ECMWF and DLWP ensembles still perform the best, and the relative performance of the various forecast systems is similar to the globally-averaged results, although the ECMWF ensemble scores are slightly improved relative to the DLWP ensemble. CRPS scores are worse in the northern hemisphere extra-tropics, 30$^{\circ}$N -- 90$^{\circ}$N (Fig.~\ref{crps}c,d), and there is a pronounced seasonality in performance.  The scores are worse in boreal winter, and the performance of the DLWP ensemble relative to the ECMWF ensemble is also worse.  On the other hand, in boreal summer, the CRPS values improve and the DWLP ensemble ties ECMWF in weeks 3, 4 and 5--6.  This seasonal difference in extra-tropical performance suggests that the DLWP model performs worse relative to ECMWF when synoptic-scale dynamics exert more influence on the weather.

\subsubsection{Ranked probability skill score\label{RPSS}}

The other metric we use to evaluate the ensemble forecasts is the ranked probability skill score (RPSS).
To compute the RPSS, $K$ categorical forecasts are first defined. Then let $y_i$ be the probabilistic forecast of the event occurring in category $i$; let $c_i$ be the climatological probability of the event falling in category $i$, and bin the verification such that $o_i=1$ if the event was observed to be in category $i$ and $o_i=0$ otherwise.  The
$k$th components of the cumulative forecast, climatological, and observational distributions $Y_k$, $C_k$, and $O_k$ are evaluated for each of the $K$ categories as $Y_k = \sum_{i=1}^k y_i$, $C_k = \sum_{i=1}^k c_i$, and $O_k = \sum_{i=1}^k o_i$.

Ranked probability scores for the forecast (RPS) and climatology (RPS$_C$) are computed as
\begin{align}
	\text{RPS} &= \sum_{k=1}^K \left(Y_k - O_k\right)^2 \label{eqn_rps} \\
	\text{RPS}_C &= \sum_{k=1}^K \left(C_k - O_k\right)^2, \label{eqn_rpsc}
\end{align}
and finally, using angled brackets to denote the average over all forecast-observation pairs, the RPSS is defined as
\begin{equation}
	\text{RPSS} = 1 - \frac{\langle \text{RPS}\rangle}{\langle \text{RPS}_C \rangle}.
\label{RPSS_def}
\end{equation}

The RPS is zero for a perfect forecast and increases positively otherwise, therefore the RPSS for a perfect forecast is one and decreases otherwise.  Normalizing $\langle \text{RPS}\rangle$  by $\langle \text{RPS}_C\rangle$ in (\ref{RPSS_def}) sets the threshold value below which there is no skill relative to climatology to zero.  Like the CRPS, the RPSS is a proper score, but it does depend on the definition of categories. 
The RPSS is sensitive to the size of the ensemble, having a negative bias for small ensemble sizes \cite{Weigel2007}. 
Although both the grand ensemble size of 320 and the ensemble size of 50 for ECMWF are large enough to mitigate such bias, in lieu of (\ref{RPSS_def}) we will use the de-biased formulation of the RPSS \cite{Weigel2007}, which for ensemble size $M$ is
\begin{align}
	\text{RPSS} &= 1 - \frac{\langle \text{RPS}\rangle}{\langle \text{RPS}_C \rangle + D_0/M} \\
	D_0 &= \frac{K^2 - 1}{6K}.
\end{align}

We compute the RPSS for $T_2$ and $T_{850}$, binning into three climatologically equally-likely terciles of below-, near-, and above-normal relative to the baseline period of 1981--2010. 
Tercile bounds for $T_2$ were determined from daily-averaged values (using the times 0, 6, 12, and 18 UTC) for each date in the 30-year record. These daily tercile bounds are then averaged over each one- or two-week verification period.
For $T_{850}$, only the instantaneous 0 UTC values were available from the ECMWF S2S database, therefore a separate climatology is computed from only 0 UTC values to evaluate the ECMWF model (the DLWP forecasts are daily-averaged and evaluated using terciles computed from the daily averages). 
Because each of the forecast categories are, by design, equally likely, the climatological forecast is simply a 33\% likelihood of occurrence in each of the categories. Note that because (\ref{eqn_rps}) and (\ref{eqn_rpsc}) use cumulative distributions, events verifying in the near-normal category have lower expected random-chance forecast error than events verifying in either the below- or above-normal categories. 

\begin{figure}
	\centering
		\includegraphics[width=0.6\textwidth, keepaspectratio]{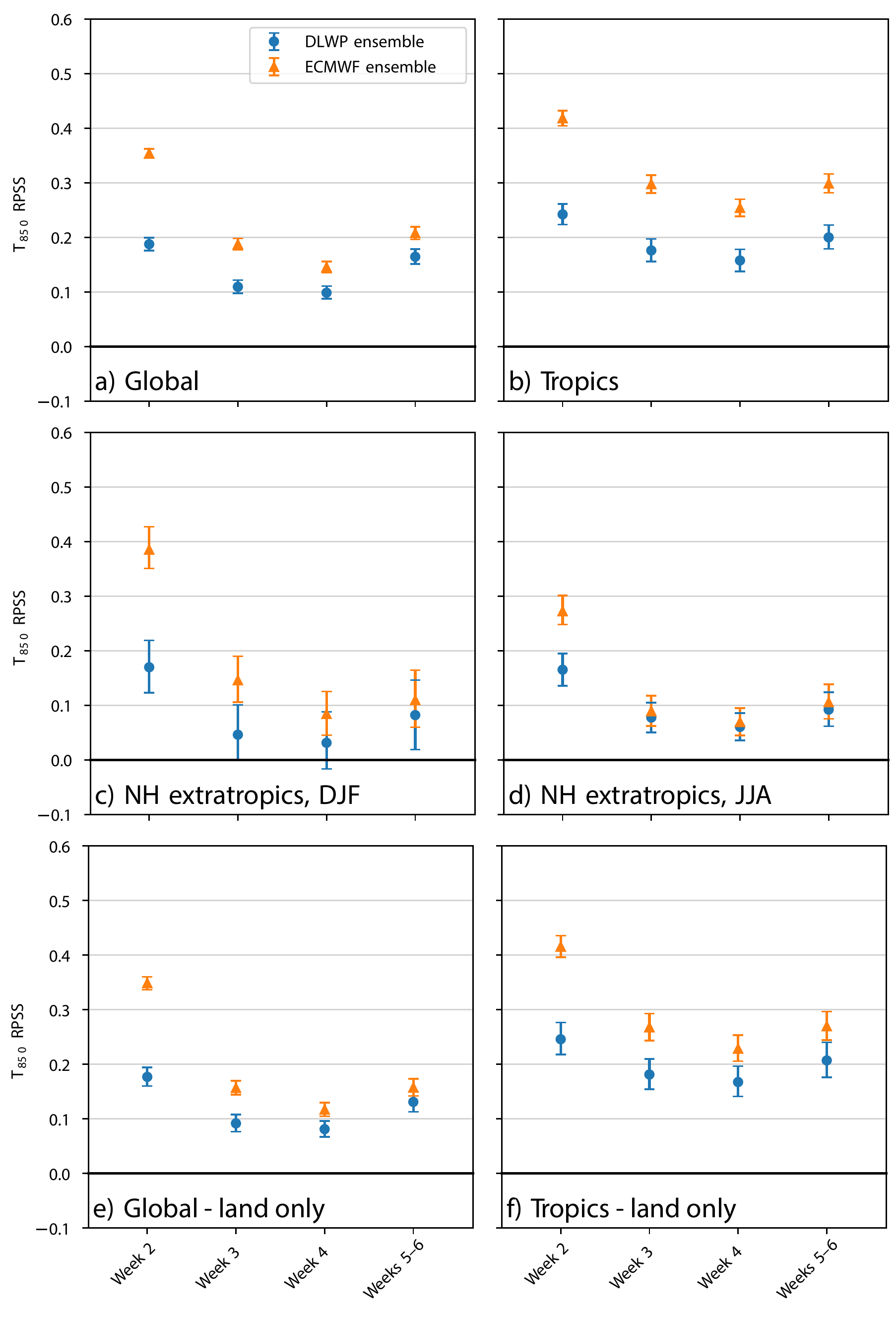}
        \caption{\label{rpss} One- or two-week averaged ranked probability skill score (RPSS; higher is better) for $T_{850}$ at indicated forecast lead times. DLWP grand ensemble (blue circles) and the ECMWF S2S ensemble (orange triangles) averaged over the (a) globe, annual mean; (b) tropics (20$^{\circ}$S -- 20$^{\circ}$N), annual mean; (c) NH extra-tropics (30$^{\circ}$N -- 90$^{\circ}$N), mean of forecasts initialized in DJF; (d) NH extra-tropics, mean over JJA; (e) and (f), as in (a) and (b) except spatially averaged only over land. Error bars correspond to the 95\% confidence interval determined by bootstrapping with 10,000 samples.}
\end{figure}

Spatially- and temporally-averaged RPSS scores in $T_{850}$ from the DLWP and ECMWF S2S ensembles are shown in Fig.~\ref{rpss}. The globally-averaged RPSS (Fig.~\ref{rpss}a) for the DLWP ensemble is well above the zero threshold for random chance at all lead times. Comparing Figs.~\ref{crps} and \ref{rpss}, and recalling that, in contrast to the RPSS, lower CRPS scores are better, both metrics show similar variations in ensemble skill with forecast lead time in all regions and over all time windows. The superiority of the ECMWF ensemble is, nevertheless, greater in the RPSS metric, with a statistically-significant lead over our DLWP ensemble at all forecast lead times and locations, except during JJA in the northern hemisphere extratropics (Fig.~\ref{rpss}d).  The performance difference between the two ensembles is significantly reduced if we consider only locations over land as shown in Fig.~\ref{rpss}e,f, where the skill of the ECMWF ensemble drops significantly after Week 2 while the RPSS for the DLWP ensemble, which has no information about SST, remains similar to that over both land and water shown in Fig.~\ref{rpss}a,b.

A similar analysis of the ECMWF ensemble RPSS, also debiased for ensemble size, was presented in \citeA{Vitart2014} for the NH extratropics (north of 30$^{\circ}$ N) and slightly different weekly periods of 12-18, 19-25 and 26-32 days.  In Table~\ref{table_rpss}, 
the performance of the DLWP and current ECMWF S2S ensembles are compared with data taken from Fig.~12a of \citeA{Vitart2014}, which gives the average performance of a five-member ensemble of reforecasts using the 2011 ECMWF forecast system in twice-weekly forecasts over the years 1995--2001.  Because sub-seasonal forecast skill can vary from year to year depending on slowly evolving large-scale circulation
patterns, caution must be used when comparing their results averaged over 1995--2001 to our calculations for 2017-2018.  Nevertheless, within the limitations of such comparison, it is worth noting that the DLWP grand ensemble performs better that the 2011 ECMWF forecast system at lead times of 19--25 and 26--32 days.

\begin{table}[t]
    \caption{\label{table_rpss} RPSS scores for $T_2$ for three S2S lead times, debiased for ensemble size, averaged globally for twice weekly forecasts in years 2017-2018 from the ECMWF and DLWP ensembles.  Column ECMWF 2011 gives the same except for reforecasts covering the years 1995--2001 using the ECMWF ensemble as formulated in 2011.}
    \centering
    \begin{tabular}{l|c c c}
     Days & ECMWF & DLWP & ECMWF 2011  \\
     \savehline
       12--18  & $0.247 \pm 0.017$ & $0.121 \pm 0.016$ & $0.135\pm 0.021$ \\
       19--25  & $0.167 \pm 0.016$ & $0.083 \pm 0.016$ & $0.054\pm 0.026$ \\
       26--32  & $0.145 \pm 0.015$ & $0.076 \pm 0.015$ & $0.030\pm 0.009$ \\
    \end{tabular}
\end{table}

\begin{figure}
	\centering
		\includegraphics[width=\textwidth, keepaspectratio]{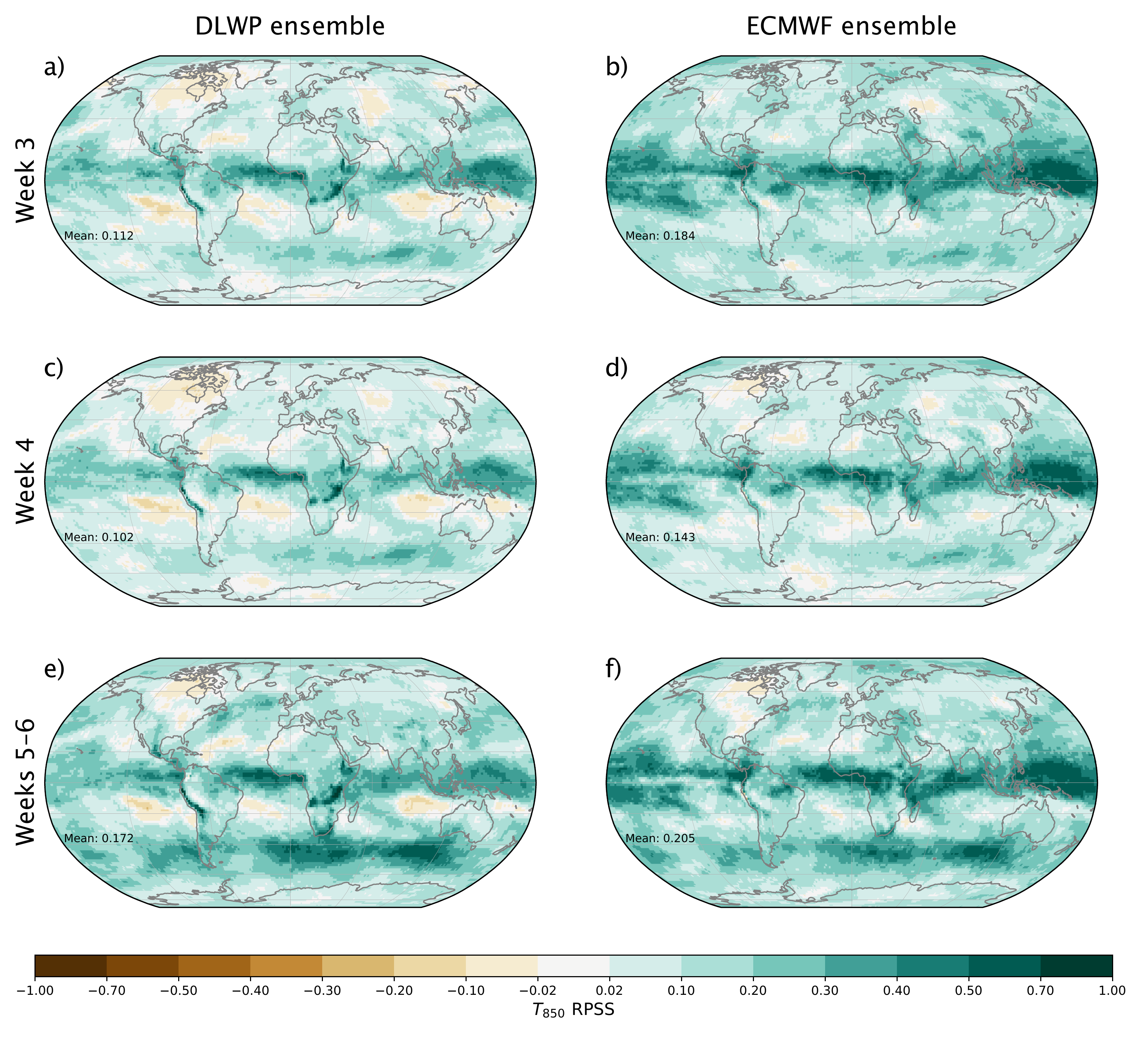}
		\caption{\label{rpss-map}  Annual average over the 2017--2018 testing period of $T_{850}$ RPSS scores. Left (right) column show DLWP (ECMWF) ensembles at forecast lead times of (a), (b): 3 weeks, (c), (d) 4 weeks, and (e), (f) 5--6 weeks.  The weighted global mean is noted in each panel.}
\end{figure}

\begin{figure}
	\centering
		\includegraphics[width=\textwidth, keepaspectratio]{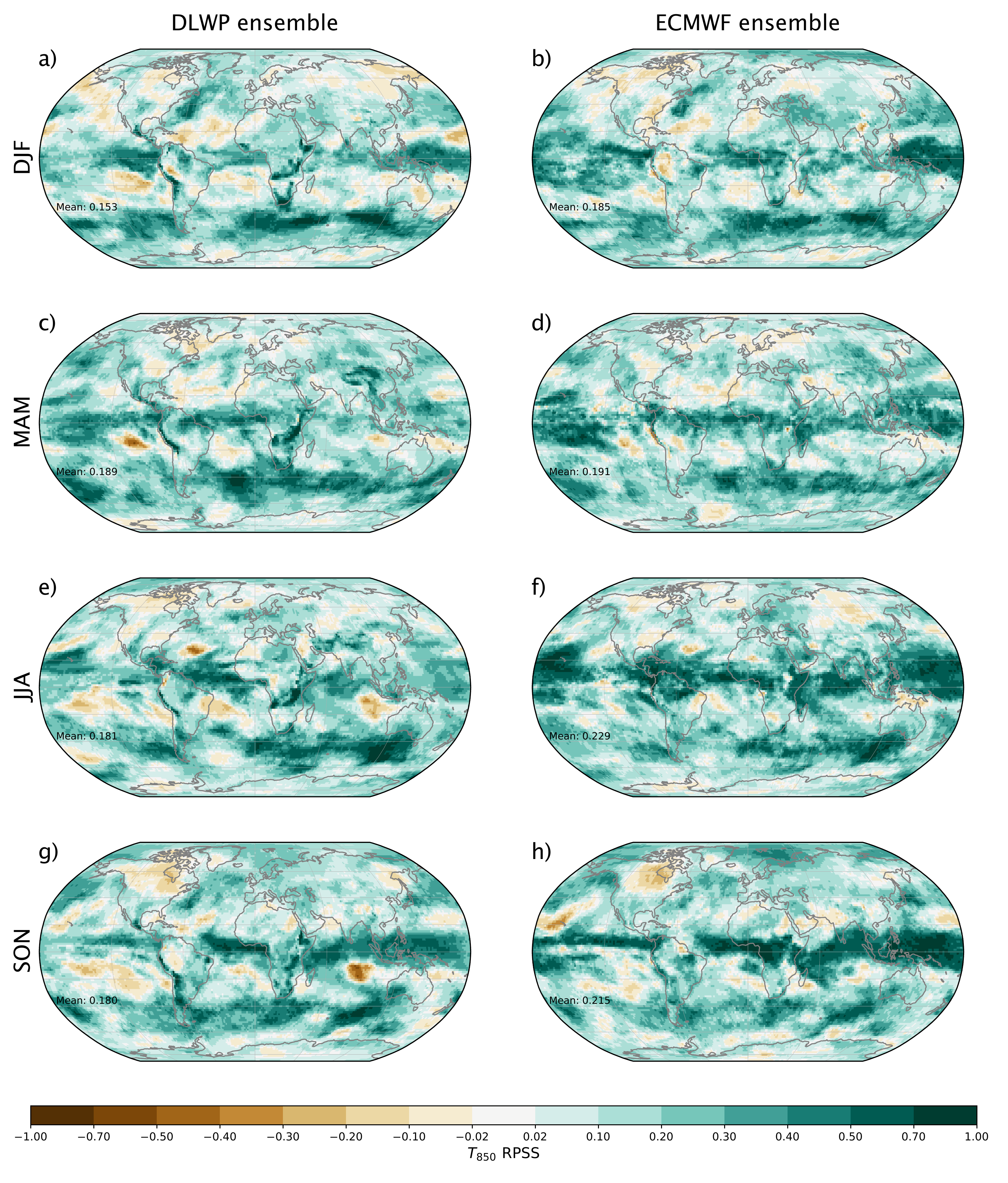}
		\caption{\label{rpss-map-season} Maps of seasonally averaged RPSS scores in $T_{850}$ during the testing period of 2017--2018. Left (right) column DLWP (ECMWF S2S) ensembles at 5--6 weeks forecast lead time for months (a), (b): DJF, (c), (d) MAM, (e), (f) JJA, and (g), (h) SON. The weighted global mean is noted in each panel.}
\end{figure}

The global pattern of RPSS scores, averaged over all of the forecasts in the 2017--2018 test set, is shown by the maps of RPSS scores for both ensembles at weeks 3, 4, and 5--6  in Fig.~\ref{rpss-map}. As expected from the plot of global average scores in Fig.~\ref{rpss}a, the ECMWF ensemble is superior to the DLWP ensemble, with the two becoming more alike and showing more skill in the forecasts averaged over the longer two-week period, weeks 5--6. Particularly at weeks 5--6 (Fig.~\ref{rpss-map}e,f), the distribution of the low- and high-skill regions in the DLWP and ECMWF ensembles are quite similar. Both ensembles perform almost the same over land and both do very well over the Southern Ocean.  The DLWP ensemble shows skill in the tropical oceans, but the ECMWF ensemble does much better in that region, and it largely avoids the loss of skill suffered by the DLWP ensemble over the adjacent subtropical waters.  As mentioned previously, our DLWP model likely suffers from the absence of information about SSTs.

The seasonal variation in RPSS at 5-6 weeks is shown in Fig.~\ref{rpss-map-season}.
The performance of the DLWP and ECMWF ensembles is most similar in MAM, with global RPSS averages of 0.180 and 0.191, respectively, and analogous spatial patterns of high and low skill.  As in the annual mean (Fig.~\ref{rpss-map}e,f), the skill is relatively high over the tropical and southern oceans. The seasonal averages show more pronounced localized regions of high and no skill ($\text{RPSS}<0$). One local area where both ensembles show skill, with the DLWP performing best, is in the storm track off the east coast of North America in both DJF and SON.
The worst globally-averaged RPSS for both ensembles, and the worst performance of DLWP relative to ECMWF, occurs during DJF, but even in this season the spatial patterns of high and low skill are similar.

\begin{figure}
	\centering
		\includegraphics[width=\textwidth, keepaspectratio]{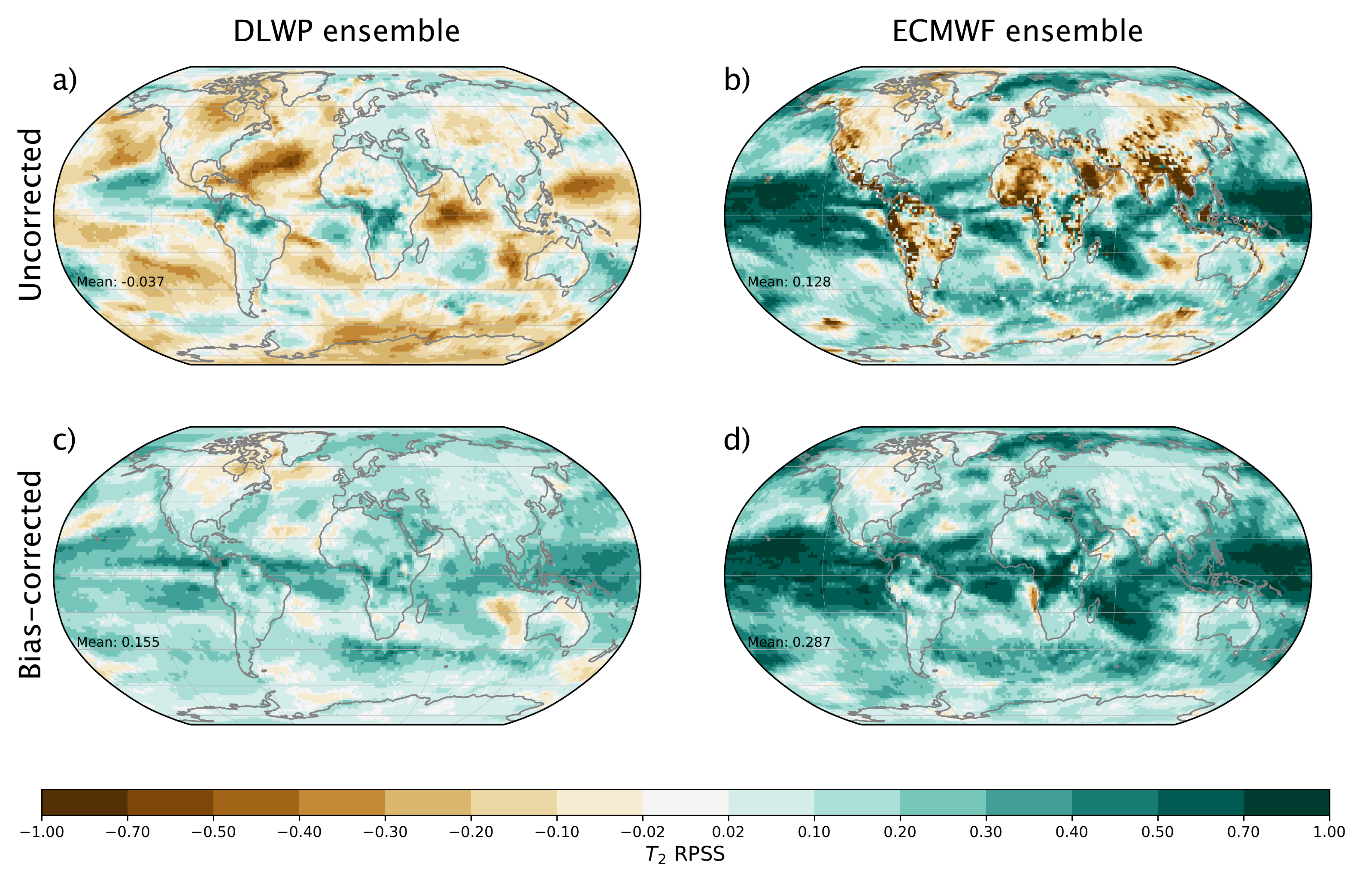}
		\caption{\label{rpss-map-t2}  Annual average RPSS skill maps for $T_2$ at weeks 5--6. Without bias correction: (a) DLWP ensemble, (b) ECMWF ensemble;  with bias correction: (c) DLWP ensemble, (d)  ECMWF ensemble. The weighted global mean is noted in each panel.}
\end{figure}

Finally, we consider surface temperature anomalies, which, as might be  expected given the pronounced model drift evident in Fig.~\ref{bias-map}, are significantly improved by bias correction.  Maps of RPSS for both the DLWP and ECMWF ensemble forecasts for weeks 5--6 are shown in Fig.~\ref{rpss-map-t2}.  Bias correction significantly improves the RPSS over land for both ensembles, with much of that improvement in the ECMWF model coming from regions with topography where removing the bias helps correct for differences in the way the orography is represented at different grid resolutions.  Bias removal also makes large improvements over the oceans in the DLWP ensemble, perhaps partly compensating for the lack of SST data.  The regions of highest skill in the bias-corrected forecasts from both ensembles are mostly over the oceans, particularly in the tropics. The global mean RPSS for both ensembles, 0.155 for DLWP and 0.287 for ECMWF, are non-negative, indicating modest skill  relative to climatology. 
The results for the bias-corrected ECMWF enssemble (Fig.~\ref{bias-map}d) are generally consistent with the distribution of RPSS  scores from an earlier version of the ECMWF ensemble in \cite{Weigel2008}, which showed skill predominantly over tropical oceans after week 2, despite accounting for bias correction in the forecasts. Improvements in the ECMWF model since \cite{Weigel2008} have, nevertheless, led to generally higher RPSS values over the oceans and much better performance over the Western Pacific and Indian Oceans, as would be apparent in a comparison of these results with their Fig.~4, although such comparison must be qualified by noting the current forecasts and those in  \cite{Weigel2008} verify in different years.

\begin{figure}
	\centering
		\includegraphics[width=\textwidth, keepaspectratio]{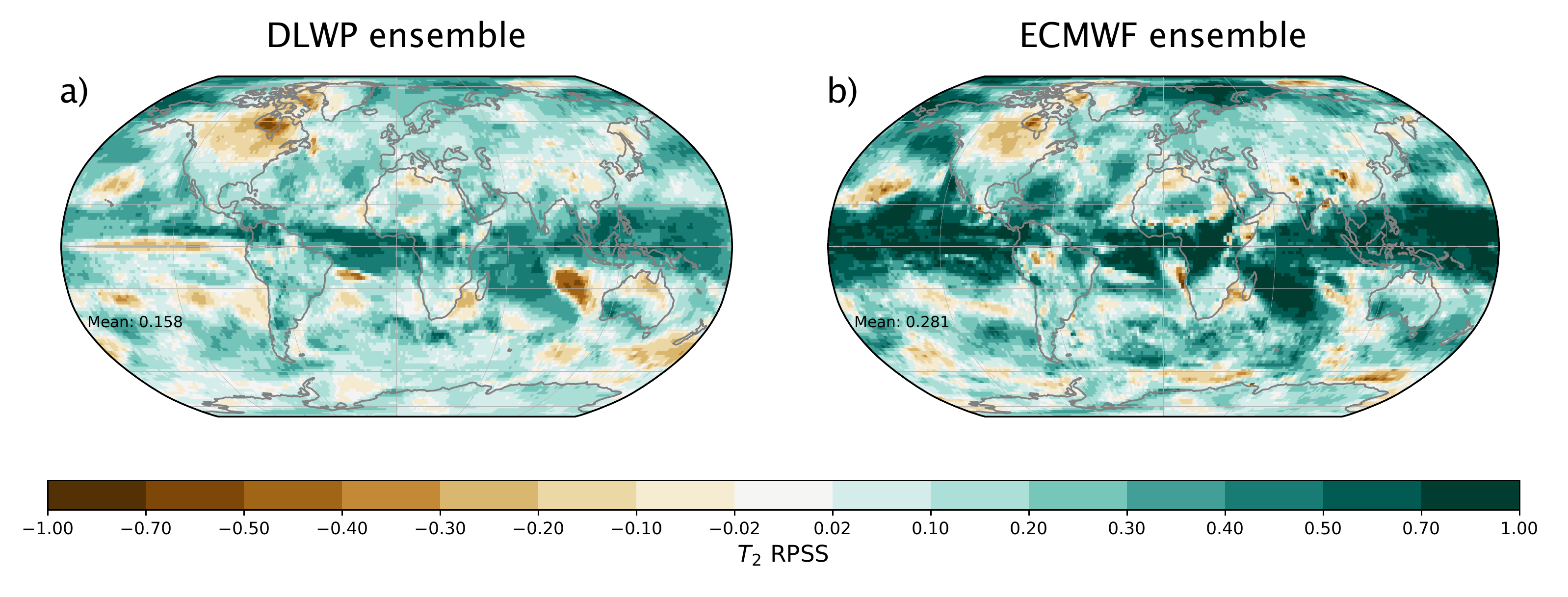}
		\caption{\label{rpss-map-t2-SON}  As in Fig.~\ref{rpss-map-t2}, but only for bias-corrected forecasts initialized in SON.}
\end{figure}

The tendency of the 5--6 week bias-corrected $T_2$ ECMWF ensemble forecast to perform similarly to the DLWP forecast over land, and better over the oceans, is again apparent in the seasonal results for SON shown in Fig.~\ref{rpss-map-t2-SON}. Note in particular, the negative RPSS score over the equatorial eastern Pacific Ocean, which arises because the DLWP ensemble fails to correctly capture the onset of a weak El Ni\~no event in 2018. In contrast, the ECMWF ensemble, with coupling to an ocean model, exhibits high skill throughout the same region.

\section{Conclusions}

As a first step toward developing a deep-learning-based ensemble system for S2S forecasting, we  refined our previous data-driven global model \cite{Weyn2020} by improving the resolution at the equator to approximately 1.4\deg by 1.4\deg and by adding two more physical fields, the temperature at 850 hPa and total column water vapor.  These refinements allowed the model to both spontaneously generate tropical cyclones and also produce a reasonable, though far from state-of-the-art, four-day deterministic forecast of hurricane Irma.  Despite the higher resolution and the expansion from four to six spherical shells of prognostic variables, the model remains very computationally efficient. A one-week forecast, stepped forward with a 12-hour time step (and 6-hour resolution), can be performed in approximately 1/10th of a second on an Nvidia Tesla V100 graphics processing unit (GPU).

We exploited this efficiency to generate large ensemble forecasts. Only about 3 minutes are required to produce a 320-member six-week ensemble forecast.  Those 320 ensemble members were generated by running 32 different DLWP models, trained to slightly different convolutional filter coefficients, on each of 10 initial conditions.  The initial conditions were non-optimal; rather than including information from singular vectors, they were simply drawn from the ERA5 archive.  The strategy of training DLWP models with slightly different filter coefficients was, on the other hand, very effective, adding significant skill to the ensemble mean and greatly increasing the ensemble spread (Figs.~\ref{ic} and \ref{ensemble}).  The ensemble spread introduced by the 32 similar DLWP models is functionally analogous to that of conventional NWP ensemble members with stochastically perturbed physical parameterization tendencies or stochastic kinetic energy backscatter.  Our DLWP model requires 6--8~days of computation to train on a single Tesla V100 GPU. We were able to economize by training only eight of our 32 ``physics-ensemble" members from scratch with different initial seeds and filling out the ensemble with models using filter coefficients from different checkpoints saved during the eight training iterations.

As was also the case for the ECMWF S2S forecast, the DLWP ensemble mean forecast was a significant improvement over that from a single control member.  In particular, the average RMSE over the 2017--2018 test set of DLWP ensemble forecasts for $Z_{500}$, $T_{850}$, and 2-m temperature remained below climatology for at least 14 days, while the anomaly correlation coefficients remained above 0.6 for 7--8 days.  Not surprisingly the ECMWF S2S ensemble did perform better, particularly at earlier lead times, and it also gave ACC scores exceeding the 0.6 threshold out to 10 days.  At longer lead times, week 3--4 or week 5-6 averages, the ACC scores of the ECMWF and DLWP ensemble means were positive and better than persistence, but still relatively low. The 2017--2018 averaged scores ranged from roughly 0.25 to 0.5, with the ECMWF ensemble performing better in all cases except for $T_{850}$ at weeks 5--6, for which both ensembles were in a statistical tie with an ACC of approximately 0.25.

We examined two probabilistic measures of ensemble skill, the CRPS and RPSS. The DLWP and ECMWF S2S ensembles produce essentially the same week-4 and weeks-5--6 CRPS scores. At shorter lead times the ECMWF ensemble is superior, performing marginally better at week 3 and distinctly better than the DLWP ensemble at week 2.  Both the DLWP and ECMWF ensembles clearly out-performed climatology and persistence.  Examining seasonal and regional contrasts showed that in the northern hemisphere extra-tropics, the DLWP ensemble performed best, and on a par with the ECMWF ensemble in the summer season; while performing worst in winter.  

Like the CRPS, the spatially- and temporally-averaged RPSS scores showed modest skill relative to climatology at all lead times. The ECMWF ensemble RPSS scores exceeded those of the DWLP ensemble by larger margins than in the CRPS metric, except in summer in the northern extra-tropics when both ensembles again achieved similar scores.  In both the globally-averaged and tropics-only-averaged RPSS, the differential by which the ECMWF RPSS score exceeds that of DLWP is smaller over land than over the full globe.  Global maps comparing the ECMWF and DLWP RPSS scores show generally similar regions of higher and lower skill, except that the ECMWF ensemble performs better over the tropical oceans. At weeks 5-6, the spatial distribution of regions of skill and no-skill in the RPSS metric over land are surprisingly similar between the ECMWF and DLWP ensembles (Fig.~\ref{rpss-map-season}). One reason the DLWP model performs poorer over the tropical oceans is likely due to its lack of SST data, as suggested by its failure in the eastern equatorial Pacific during the onset of a weak El Ni\~no event in 2018. 

Although our current data-driven DLWP model is worse than operational NWP models for the deterministic prediction of synoptic-scale weather patterns, its capability to learn physics-based phenomena, including the complex evolution of near-surface temperatures and long-term patterns in the convection-dominated tropics, is remarkable.  As such, DLWP may prove a valuable tool for supplementing NWP-based S2S forecasts where they are weakest: in the tropics and in the spring and summer months. 

There are many avenues for further development of our elementary DLWP ensemble system. One obvious shortcoming is that our DLWP model does not yet forecast precipitation. This might be addressed by adding precipitation to the current set of six prognostic 2D fields that are recursively stepped forward by the model, although instead of including the precipitation at previous times in the CNN architecture, it could alternatively be diagnosed from the other fields after each step \cite{Larraondo2019}. 

The DLWP model's computational efficiency can be used for more than simply producing timely operational forecasts; it also enables researchers to make unprecedented use of large numbers of reforecasts for past weather events. 
We computed 25~years of reforecasts, with 104 forecasts per year for 33 ensemble members, in a matter of hours on a single GPU. We only used these reforecasts to correct the average drift in the DLWP model, but one could also use them to calibrate ensemble probability distributions, analyze model errors, or investigate of the sources of predictability captured by the model. 
Historically, adjoint models \cite<e.g.,>{Doyle2014}, which are tangent linear, differentiable approximations to full non-linear dynamical NWP models, have been used to examine how model errors depend on initial condition uncertainties (for example, whether errors in moisture over the US have a strong influence on the location or intensity of a cyclone over Europe). 
Adjoint models are difficult to create for complex operational NWP models. Yet, because a CNN is fully differentiable, it is easy to produce the corresponding adjoint model, enabling studies of error growth and atmospheric predictability. 
Recent work on the interpretation of deep neural networks may provide some valuable tools for this form of analysis \cite{Toms2020,Ebert-Uphoff2020}.

\acknowledgments
J. A. Weyn and D. R. Durran's contributions to this research were funded by Grant N00014-17-1-2660 from the Office of Naval Research (ONR). D. R. Durran was also supported by Grant N00014-20-1-2387 from ONR. J. A. Weyn was also supported by a National Defense Science and Engineering Graduate (NDSEG) fellowship from the Department of Defense (DoD). Computational resources were provided by Microsoft Azure via a grant from Microsoft's AI for Earth program. Stan Posey (Nvidia) also generously donated two V100 GPUs to the University of Washington which were used for this work.

\datastatement
The ERA5 reanalysis data are available via the Copernicus Climate Data Store (DOI: 10.24381/cds.bd0915c6 and DOI: 10.24381/cds.adbb2d47). The ECMWF S2S forecasts are available
at https://apps.ecmwf.int/datasets/data/s2s. The T42 and T63 IFS forecasts are available from Rasp et al.
(2020b).


%
%

\bibliography{dlwpS2S.bib}

\begin{thebibliography}{}

\bibitem [\protect \citeauthoryear {%
Adames%
\ \BBA {} Kim%
}{%
Adames%
\ \BBA {} Kim%
}{%
{\protect \APACyear {2016}}%
}]{%
adames2016}
\APACinsertmetastar {%
adames2016}%
\begin{APACrefauthors}%
Adames, {\'A}\BPBI F.%
\BCBT {}\ \BBA {} Kim, D.%
\end{APACrefauthors}%
\unskip\
\newblock
\APACrefYearMonthDay{2016}{{\APACmonth{03}}}{}.
\newblock
{\BBOQ}\APACrefatitle {The {MJO} as a Dispersive, Convectively Coupled Moisture
  Wave: Theory and Observations} {The {MJO} as a dispersive, convectively
  coupled moisture wave: Theory and observations}.{\BBCQ}
\newblock
\APACjournalVolNumPages{Journal of the Atmospheric Sciences}{73}{3}{913--941}.
\newblock
\begin{APACrefDOI} \doi{10.1175/JAS-D-15-0170.1} \end{APACrefDOI}
\PrintBackRefs{\CurrentBib}

\bibitem [\protect \citeauthoryear {%
Bauer%
, Thorpe%
\BCBL {}\ \BBA {} Brunet%
}{%
Bauer%
\ \protect \BOthers {.}}{%
{\protect \APACyear {2015}}%
}]{%
Bauer2015}
\APACinsertmetastar {%
Bauer2015}%
\begin{APACrefauthors}%
Bauer, P.%
, Thorpe, A.%
\BCBL {}\ \BBA {} Brunet, G.%
\end{APACrefauthors}%
\unskip\
\newblock
\APACrefYearMonthDay{2015}{}{}.
\newblock
{\BBOQ}\APACrefatitle {{The quiet revolution of numerical weather prediction}}
  {{The quiet revolution of numerical weather prediction}}.{\BBCQ}
\newblock
\APACjournalVolNumPages{Nature}{525}{7567}{}.
\newblock
\begin{APACrefDOI} \doi{10.1038/nature14956} \end{APACrefDOI}
\PrintBackRefs{\CurrentBib}

\bibitem [\protect \citeauthoryear {%
Buizza%
}{%
Buizza%
}{%
{\protect \APACyear {2019}}%
}]{%
Buizza2019}
\APACinsertmetastar {%
Buizza2019}%
\begin{APACrefauthors}%
Buizza, R.%
\end{APACrefauthors}%
\unskip\
\newblock
\APACrefYearMonthDay{2019}{}{}.
\newblock
{\BBOQ}\APACrefatitle {Introduction to the special issue on ``25 years of
  ensemble forecasting"} {Introduction to the special issue on ``25 years of
  ensemble forecasting"}.{\BBCQ}
\newblock
\APACjournalVolNumPages{Quarterly Journal of the Royal Meteorological
  Society}{145}{S1}{1--11}.
\newblock
\begin{APACrefDOI} \doi{10.1002/qj.3370} \end{APACrefDOI}
\PrintBackRefs{\CurrentBib}

\bibitem [\protect \citeauthoryear {%
Doyle%
, Amerault%
, Reynolds%
\BCBL {}\ \BBA {} Reinecke%
}{%
Doyle%
\ \protect \BOthers {.}}{%
{\protect \APACyear {2014}}%
}]{%
Doyle2014}
\APACinsertmetastar {%
Doyle2014}%
\begin{APACrefauthors}%
Doyle, J\BPBI D.%
, Amerault, C.%
, Reynolds, C\BPBI A.%
\BCBL {}\ \BBA {} Reinecke, P\BPBI A.%
\end{APACrefauthors}%
\unskip\
\newblock
\APACrefYearMonthDay{2014}{}{}.
\newblock
{\BBOQ}\APACrefatitle {{Initial condition sensitivity and predictability of a
  severe extratropical cyclone using a moist adjoint}} {{Initial condition
  sensitivity and predictability of a severe extratropical cyclone using a
  moist adjoint}}.{\BBCQ}
\newblock
\APACjournalVolNumPages{Monthly Weather Review}{142}{1}{320--342}.
\newblock
\begin{APACrefDOI} \doi{10.1175/MWR-D-13-00201.1} \end{APACrefDOI}
\PrintBackRefs{\CurrentBib}

\bibitem [\protect \citeauthoryear {%
Dueben%
\ \BBA {} Bauer%
}{%
Dueben%
\ \BBA {} Bauer%
}{%
{\protect \APACyear {2018}}%
}]{%
Dueben2018}
\APACinsertmetastar {%
Dueben2018}%
\begin{APACrefauthors}%
Dueben, P\BPBI D.%
\BCBT {}\ \BBA {} Bauer, P.%
\end{APACrefauthors}%
\unskip\
\newblock
\APACrefYearMonthDay{2018}{}{}.
\newblock
{\BBOQ}\APACrefatitle {{Challenges and design choices for global weather and
  climate models based on machine learning}} {{Challenges and design choices
  for global weather and climate models based on machine learning}}.{\BBCQ}
\newblock
\APACjournalVolNumPages{Geoscientific Model Development}{11}{10}{3999--4009}.
\newblock
\begin{APACrefDOI} \doi{10.5194/gmd-11-3999-2018} \end{APACrefDOI}
\PrintBackRefs{\CurrentBib}

\bibitem [\protect \citeauthoryear {%
Ebert-Uphoff%
\ \BBA {} Hilburn%
}{%
Ebert-Uphoff%
\ \BBA {} Hilburn%
}{%
{\protect \APACyear {2020}}%
}]{%
Ebert-Uphoff2020}
\APACinsertmetastar {%
Ebert-Uphoff2020}%
\begin{APACrefauthors}%
Ebert-Uphoff, I.%
\BCBT {}\ \BBA {} Hilburn, K\BPBI A.%
\end{APACrefauthors}%
\unskip\
\newblock
\APACrefYearMonthDay{2020}{may}{}.
\newblock
{\BBOQ}\APACrefatitle {{Evaluation, Tuning and Interpretation of Neural
  Networks for Meteorological Applications}} {{Evaluation, Tuning and
  Interpretation of Neural Networks for Meteorological Applications}}.{\BBCQ}
\newblock
\APACjournalVolNumPages{ArXiv}{}{}{}.
\newblock
\begin{APACrefURL} \url{http://arxiv.org/abs/2005.03126} \end{APACrefURL}
\PrintBackRefs{\CurrentBib}

\bibitem [\protect \citeauthoryear {%
Fortin%
, Abaza%
, Anctil%
\BCBL {}\ \BBA {} Turcotte%
}{%
Fortin%
\ \protect \BOthers {.}}{%
{\protect \APACyear {2014}}%
}]{%
Fortin2014}
\APACinsertmetastar {%
Fortin2014}%
\begin{APACrefauthors}%
Fortin, V.%
, Abaza, M.%
, Anctil, F.%
\BCBL {}\ \BBA {} Turcotte, R.%
\end{APACrefauthors}%
\unskip\
\newblock
\APACrefYearMonthDay{2014}{Aug}{}.
\newblock
{\BBOQ}\APACrefatitle {Why Should Ensemble Spread Match the RMSE of the
  Ensemble Mean?} {Why should ensemble spread match the rmse of the ensemble
  mean?}{\BBCQ}
\newblock
\APACjournalVolNumPages{J.~Hydrometeorol.}{15}{4}{1708–1713}.
\newblock
\begin{APACrefDOI} \doi{10.1175/JHM-D-14-0008.1} \end{APACrefDOI}
\PrintBackRefs{\CurrentBib}

\bibitem [\protect \citeauthoryear {%
Gneiting%
\ \BBA {} Raftery%
}{%
Gneiting%
\ \BBA {} Raftery%
}{%
{\protect \APACyear {2007}}%
}]{%
Gneiting2007}
\APACinsertmetastar {%
Gneiting2007}%
\begin{APACrefauthors}%
Gneiting, T.%
\BCBT {}\ \BBA {} Raftery, A\BPBI E.%
\end{APACrefauthors}%
\unskip\
\newblock
\APACrefYearMonthDay{2007}{}{}.
\newblock
{\BBOQ}\APACrefatitle {{Strictly proper scoring rules, prediction, and
  estimation}} {{Strictly proper scoring rules, prediction, and
  estimation}}.{\BBCQ}
\newblock
\APACjournalVolNumPages{Journal of the American Statistical
  Association}{102}{477}{359--378}.
\newblock
\begin{APACrefDOI} \doi{10.1198/016214506000001437} \end{APACrefDOI}
\PrintBackRefs{\CurrentBib}

\bibitem [\protect \citeauthoryear {%
Hersbach%
}{%
Hersbach%
}{%
{\protect \APACyear {2000}}%
}]{%
Hersbach2000}
\APACinsertmetastar {%
Hersbach2000}%
\begin{APACrefauthors}%
Hersbach, H.%
\end{APACrefauthors}%
\unskip\
\newblock
\APACrefYearMonthDay{2000}{}{}.
\newblock
{\BBOQ}\APACrefatitle {{Decomposition of the continuous ranked probability
  score for ensemble prediction systems}} {{Decomposition of the continuous
  ranked probability score for ensemble prediction systems}}.{\BBCQ}
\newblock
\APACjournalVolNumPages{Weather and Forecasting}{15}{5}{559--570}.
\newblock
\begin{APACrefDOI} \doi{10.1175/1520-0434(2000)015<0559:DOTCRP>2.0.CO;2}
  \end{APACrefDOI}
\PrintBackRefs{\CurrentBib}

\bibitem [\protect \citeauthoryear {%
Hwang%
, Orenstein%
, Cohen%
, Pfeiffer%
\BCBL {}\ \BBA {} Mackey%
}{%
Hwang%
\ \protect \BOthers {.}}{%
{\protect \APACyear {2019}}%
}]{%
Hwang2019}
\APACinsertmetastar {%
Hwang2019}%
\begin{APACrefauthors}%
Hwang, J.%
, Orenstein, P.%
, Cohen, J.%
, Pfeiffer, K.%
\BCBL {}\ \BBA {} Mackey, L.%
\end{APACrefauthors}%
\unskip\
\newblock
\APACrefYearMonthDay{2019}{sep}{}.
\newblock
{\BBOQ}\APACrefatitle {{Improving subseasonal forecasting in the western U.S.
  With machine learning}} {{Improving subseasonal forecasting in the western
  U.S. With machine learning}}.{\BBCQ}
\newblock
\BIn{} \APACrefbtitle {Proceedings of the ACM SIGKDD International Conference
  on Knowledge Discovery and Data Mining} {Proceedings of the acm sigkdd
  international conference on knowledge discovery and data mining}\ (\BPGS\
  2325--2335).
\newblock
\begin{APACrefURL} \url{http://arxiv.org/abs/1809.07394
  https://arxiv.org/abs/1809.07394} \end{APACrefURL}
\newblock
\begin{APACrefDOI} \doi{10.1145/3292500.3330674} \end{APACrefDOI}
\PrintBackRefs{\CurrentBib}

\bibitem [\protect \citeauthoryear {%
Isaksen%
\ \protect \BOthers {.}}{%
Isaksen%
\ \protect \BOthers {.}}{%
{\protect \APACyear {2010}}%
}]{%
Isaksen2010}
\APACinsertmetastar {%
Isaksen2010}%
\begin{APACrefauthors}%
Isaksen, L.%
, Bonavita, M.%
, Buizza, R.%
, Fisher, M.%
, Haseler, J.%
, Leutbecher, M.%
\BCBL {}\ \BBA {} Raynaud, L.%
\end{APACrefauthors}%
\unskip\
\newblock
\APACrefYearMonthDay{2010}{}{}.
\newblock
{\BBOQ}\APACrefatitle {{Ensemble of data assimilations at ECMWF}} {{Ensemble of
  data assimilations at ECMWF}}.{\BBCQ}
\newblock
\APACjournalVolNumPages{ECMWF Tech. Memo.}{636}{December}{1--41}.
\PrintBackRefs{\CurrentBib}

\bibitem [\protect \citeauthoryear {%
Kingma%
\ \BBA {} Ba%
}{%
Kingma%
\ \BBA {} Ba%
}{%
{\protect \APACyear {2014}}%
}]{%
Kingma2014}
\APACinsertmetastar {%
Kingma2014}%
\begin{APACrefauthors}%
Kingma, D\BPBI P.%
\BCBT {}\ \BBA {} Ba, J.%
\end{APACrefauthors}%
\unskip\
\newblock
\APACrefYearMonthDay{2014}{}{}.
\newblock
{\BBOQ}\APACrefatitle {{Adam: A Method for Stochastic Optimization}} {{Adam: A
  Method for Stochastic Optimization}}.{\BBCQ}
\newblock
\APACjournalVolNumPages{ArXiv}{}{}{}.
\newblock
\begin{APACrefURL} \url{http://arxiv.org/abs/1412.6980} \end{APACrefURL}
\PrintBackRefs{\CurrentBib}

\bibitem [\protect \citeauthoryear {%
Larraondo%
, Renzullo%
, Inza%
\BCBL {}\ \BBA {} Lozano%
}{%
Larraondo%
\ \protect \BOthers {.}}{%
{\protect \APACyear {2019}}%
}]{%
Larraondo2019}
\APACinsertmetastar {%
Larraondo2019}%
\begin{APACrefauthors}%
Larraondo, P\BPBI R.%
, Renzullo, L\BPBI J.%
, Inza, I.%
\BCBL {}\ \BBA {} Lozano, J\BPBI A.%
\end{APACrefauthors}%
\unskip\
\newblock
\APACrefYearMonthDay{2019}{}{}.
\newblock
{\BBOQ}\APACrefatitle {{A data-driven approach to precipitation
  parameterizations using convolutional encoder-decoder neural networks}} {{A
  data-driven approach to precipitation parameterizations using convolutional
  encoder-decoder neural networks}}.{\BBCQ}
\newblock
\APACjournalVolNumPages{ArXiv}{}{}{}.
\newblock
\begin{APACrefURL} \url{http://arxiv.org/abs/1903.10274} \end{APACrefURL}
\PrintBackRefs{\CurrentBib}

\bibitem [\protect \citeauthoryear {%
Leutbecher%
}{%
Leutbecher%
}{%
{\protect \APACyear {2018}}%
}]{%
Leutbecher2018}
\APACinsertmetastar {%
Leutbecher2018}%
\begin{APACrefauthors}%
Leutbecher, M.%
\end{APACrefauthors}%
\unskip\
\newblock
\APACrefYearMonthDay{2018}{sep}{}.
\newblock
{\BBOQ}\APACrefatitle {{Ensemble size: How suboptimal is less than infinity?}}
  {{Ensemble size: How suboptimal is less than infinity?}}{\BBCQ}
\newblock
\APACjournalVolNumPages{Quarterly Journal of the Royal Meteorological
  Society}{145}{S1}{107--128}.
\newblock
\begin{APACrefDOI} \doi{10.1002/qj.3387} \end{APACrefDOI}
\PrintBackRefs{\CurrentBib}

\bibitem [\protect \citeauthoryear {%
Monhart%
\ \protect \BOthers {.}}{%
Monhart%
\ \protect \BOthers {.}}{%
{\protect \APACyear {2018}}%
}]{%
Monhart2018}
\APACinsertmetastar {%
Monhart2018}%
\begin{APACrefauthors}%
Monhart, S.%
, Spirig, C.%
, Bhend, J.%
, Bogner, K.%
, Schär, C.%
\BCBL {}\ \BBA {} Liniger, M\BPBI A.%
\end{APACrefauthors}%
\unskip\
\newblock
\APACrefYearMonthDay{2018}{Aug}{}.
\newblock
{\BBOQ}\APACrefatitle {Skill of Subseasonal Forecasts in {E}urope: Effect of
  Bias Correction and Downscaling Using Surface Observations} {Skill of
  subseasonal forecasts in {E}urope: Effect of bias correction and downscaling
  using surface observations}.{\BBCQ}
\newblock
\APACjournalVolNumPages{Journal of Geophysical Research: Atmospheres}{}{}{}.
\newblock
\begin{APACrefURL} \url{http://doi.wiley.com/10.1029/2017JD027923}
  \end{APACrefURL}
\newblock
\begin{APACrefDOI} \doi{10.1029/2017JD027923} \end{APACrefDOI}
\PrintBackRefs{\CurrentBib}

\bibitem [\protect \citeauthoryear {%
Palmer%
}{%
Palmer%
}{%
{\protect \APACyear {2018}}%
}]{%
Palmer2018}
\APACinsertmetastar {%
Palmer2018}%
\begin{APACrefauthors}%
Palmer, T.%
\end{APACrefauthors}%
\unskip\
\newblock
\APACrefYearMonthDay{2018}{}{}.
\newblock
{\BBOQ}\APACrefatitle {{The ECMWF ensemble prediction system: Looking back
  (more than) 25 years and projecting forward 25 years}} {{The ECMWF ensemble
  prediction system: Looking back (more than) 25 years and projecting forward
  25 years}}.{\BBCQ}
\newblock
\APACjournalVolNumPages{Quarterly Journal of the Royal Meteorological
  Society}{}{}{}.
\newblock
\begin{APACrefURL} \url{http://doi.wiley.com/10.1002/qj.3383} \end{APACrefURL}
\newblock
\begin{APACrefDOI} \doi{10.1002/qj.3383} \end{APACrefDOI}
\PrintBackRefs{\CurrentBib}

\bibitem [\protect \citeauthoryear {%
Richardson%
}{%
Richardson%
}{%
{\protect \APACyear {2000}}%
}]{%
Richardson2000}
\APACinsertmetastar {%
Richardson2000}%
\begin{APACrefauthors}%
Richardson, D\BPBI S.%
\end{APACrefauthors}%
\unskip\
\newblock
\APACrefYearMonthDay{2000}{}{}.
\newblock
{\BBOQ}\APACrefatitle {Skill and relative economic value of the {ECMWF}
  ensemble prediction system} {Skill and relative economic value of the {ECMWF}
  ensemble prediction system}.{\BBCQ}
\newblock
\APACjournalVolNumPages{Quarterly Journal of the Royal Meteorological
  Society}{126}{563}{649--667}.
\newblock
\begin{APACrefDOI} \doi{https://doi.org/10.1002/qj.49712656313}
  \end{APACrefDOI}
\PrintBackRefs{\CurrentBib}

\bibitem [\protect \citeauthoryear {%
Ronneberger%
, Fischer%
\BCBL {}\ \BBA {} Brox%
}{%
Ronneberger%
\ \protect \BOthers {.}}{%
{\protect \APACyear {2015}}%
}]{%
Ronneberger2015}
\APACinsertmetastar {%
Ronneberger2015}%
\begin{APACrefauthors}%
Ronneberger, O.%
, Fischer, P.%
\BCBL {}\ \BBA {} Brox, T.%
\end{APACrefauthors}%
\unskip\
\newblock
\APACrefYearMonthDay{2015}{}{}.
\newblock
{\BBOQ}\APACrefatitle {{U-net: Convolutional networks for biomedical image
  segmentation}} {{U-net: Convolutional networks for biomedical image
  segmentation}}.{\BBCQ}
\newblock
\BIn{} \APACrefbtitle {Lecture Notes in Computer Science (including subseries
  Lecture Notes in Artificial Intelligence and Lecture Notes in
  Bioinformatics)} {Lecture notes in computer science (including subseries
  lecture notes in artificial intelligence and lecture notes in
  bioinformatics)}\ (\BVOL\ 9351, \BPGS\ 234--241).
\newblock
\APACaddressPublisher{}{Springer, Cham}.
\newblock
\begin{APACrefURL}
  \url{http://link.springer.com/10.1007/978-3-319-24574-4{\_}28}
  \end{APACrefURL}
\newblock
\begin{APACrefDOI} \doi{10.1007/978-3-319-24574-4_28} \end{APACrefDOI}
\PrintBackRefs{\CurrentBib}

\bibitem [\protect \citeauthoryear {%
Scher%
\ \BBA {} Messori%
}{%
Scher%
\ \BBA {} Messori%
}{%
{\protect \APACyear {2019}}%
}]{%
Scher2019}
\APACinsertmetastar {%
Scher2019}%
\begin{APACrefauthors}%
Scher, S.%
\BCBT {}\ \BBA {} Messori, G.%
\end{APACrefauthors}%
\unskip\
\newblock
\APACrefYearMonthDay{2019}{}{}.
\newblock
{\BBOQ}\APACrefatitle {{Weather and climate forecasting with neural networks:
  using GCMs with different complexity as study-ground}} {{Weather and climate
  forecasting with neural networks: using GCMs with different complexity as
  study-ground}}.{\BBCQ}
\newblock
\APACjournalVolNumPages{Geoscientific Model Development
  Discussions}{}{March}{1--15}.
\newblock
\begin{APACrefURL}
  \url{https://www.geosci-model-dev-discuss.net/gmd-2019-53/{\%}0Ahttps://www.geosci-model-dev-discuss.net/gmd-2019-53/}
  \end{APACrefURL}
\newblock
\begin{APACrefDOI} \doi{10.5194/gmd-2019-53} \end{APACrefDOI}
\PrintBackRefs{\CurrentBib}

\bibitem [\protect \citeauthoryear {%
Scher%
\ \BBA {} Messori%
}{%
Scher%
\ \BBA {} Messori%
}{%
{\protect \APACyear {2020}}%
}]{%
Scher2020}
\APACinsertmetastar {%
Scher2020}%
\begin{APACrefauthors}%
Scher, S.%
\BCBT {}\ \BBA {} Messori, G.%
\end{APACrefauthors}%
\unskip\
\newblock
\APACrefYearMonthDay{2020}{feb}{}.
\newblock
{\BBOQ}\APACrefatitle {{Ensemble neural network forecasts with singular value
  decomposition}} {{Ensemble neural network forecasts with singular value
  decomposition}}.{\BBCQ}
\newblock
\APACjournalVolNumPages{ArXiv}{}{}{}.
\newblock
\begin{APACrefURL} \url{http://arxiv.org/abs/2002.05398} \end{APACrefURL}
\PrintBackRefs{\CurrentBib}

\bibitem [\protect \citeauthoryear {%
Toms%
, Barnes%
\BCBL {}\ \BBA {} Ebert-Uphoff%
}{%
Toms%
\ \protect \BOthers {.}}{%
{\protect \APACyear {2020}}%
}]{%
Toms2020}
\APACinsertmetastar {%
Toms2020}%
\begin{APACrefauthors}%
Toms, B\BPBI A.%
, Barnes, E\BPBI A.%
\BCBL {}\ \BBA {} Ebert-Uphoff, I.%
\end{APACrefauthors}%
\unskip\
\newblock
\APACrefYearMonthDay{2020}{}{}.
\newblock
{\BBOQ}\APACrefatitle {{Physically Interpretable Neural Networks for the
  Geosciences: Applications to Earth System Variability}} {{Physically
  Interpretable Neural Networks for the Geosciences: Applications to Earth
  System Variability}}.{\BBCQ}
\newblock
\APACjournalVolNumPages{Journal of Advances in Modeling Earth
  Systems}{12}{9}{e2019MS002002}.
\newblock
\begin{APACrefDOI} \doi{10.1029/2019MS002002} \end{APACrefDOI}
\PrintBackRefs{\CurrentBib}

\bibitem [\protect \citeauthoryear {%
Ullrich%
, Devendran%
\BCBL {}\ \BBA {} Johansen%
}{%
Ullrich%
\ \protect \BOthers {.}}{%
{\protect \APACyear {2016}}%
}]{%
Ullrich2016}
\APACinsertmetastar {%
Ullrich2016}%
\begin{APACrefauthors}%
Ullrich, P\BPBI A.%
, Devendran, D.%
\BCBL {}\ \BBA {} Johansen, H.%
\end{APACrefauthors}%
\unskip\
\newblock
\APACrefYearMonthDay{2016}{}{}.
\newblock
{\BBOQ}\APACrefatitle {{Arbitrary-order conservative and consistent remapping
  and a theory of linear maps: Part II}} {{Arbitrary-order conservative and
  consistent remapping and a theory of linear maps: Part II}}.{\BBCQ}
\newblock
\APACjournalVolNumPages{Monthly Weather Review}{144}{4}{1529--1549}.
\newblock
\begin{APACrefDOI} \doi{10.1175/MWR-D-15-0301.1} \end{APACrefDOI}
\PrintBackRefs{\CurrentBib}

\bibitem [\protect \citeauthoryear {%
Ullrich%
\ \BBA {} Taylor%
}{%
Ullrich%
\ \BBA {} Taylor%
}{%
{\protect \APACyear {2015}}%
}]{%
Ullrich2015}
\APACinsertmetastar {%
Ullrich2015}%
\begin{APACrefauthors}%
Ullrich, P\BPBI A.%
\BCBT {}\ \BBA {} Taylor, M\BPBI A.%
\end{APACrefauthors}%
\unskip\
\newblock
\APACrefYearMonthDay{2015}{}{}.
\newblock
{\BBOQ}\APACrefatitle {{Arbitrary-order conservative and consistent remapping
  and a theory of linear maps: Part I}} {{Arbitrary-order conservative and
  consistent remapping and a theory of linear maps: Part I}}.{\BBCQ}
\newblock
\APACjournalVolNumPages{Monthly Weather Review}{143}{6}{2419--2440}.
\newblock
\begin{APACrefDOI} \doi{10.1175/MWR-D-14-00343.1} \end{APACrefDOI}
\PrintBackRefs{\CurrentBib}

\bibitem [\protect \citeauthoryear {%
Vitart%
}{%
Vitart%
}{%
{\protect \APACyear {2004}}%
}]{%
Vitart2004}
\APACinsertmetastar {%
Vitart2004}%
\begin{APACrefauthors}%
Vitart, F.%
\end{APACrefauthors}%
\unskip\
\newblock
\APACrefYearMonthDay{2004}{}{}.
\newblock
{\BBOQ}\APACrefatitle {{Monthly forecasting at ECMWF}} {{Monthly forecasting at
  ECMWF}}.{\BBCQ}
\newblock
\APACjournalVolNumPages{Monthly Weather Review}{132}{12}{2761--2779}.
\newblock
\begin{APACrefDOI} \doi{10.1175/MWR2826.1} \end{APACrefDOI}
\PrintBackRefs{\CurrentBib}

\bibitem [\protect \citeauthoryear {%
Vitart%
}{%
Vitart%
}{%
{\protect \APACyear {2014}}%
}]{%
Vitart2014}
\APACinsertmetastar {%
Vitart2014}%
\begin{APACrefauthors}%
Vitart, F.%
\end{APACrefauthors}%
\unskip\
\newblock
\APACrefYearMonthDay{2014}{}{}.
\newblock
{\BBOQ}\APACrefatitle {{Evolution of ECMWF sub-seasonal forecast skill scores}}
  {{Evolution of ECMWF sub-seasonal forecast skill scores}}.{\BBCQ}
\newblock
\APACjournalVolNumPages{Quarterly Journal of the Royal Meteorological
  Society}{140}{683}{1889--1899}.
\newblock
\begin{APACrefDOI} \doi{10.1002/qj.2256} \end{APACrefDOI}
\PrintBackRefs{\CurrentBib}

\bibitem [\protect \citeauthoryear {%
Vitart%
\ \protect \BOthers {.}}{%
Vitart%
\ \protect \BOthers {.}}{%
{\protect \APACyear {2017}}%
}]{%
Vitart2017}
\APACinsertmetastar {%
Vitart2017}%
\begin{APACrefauthors}%
Vitart, F.%
, Ardilouze, C.%
, Bonet, A.%
, Brookshaw, A.%
, Chen, M.%
, Codorean, C.%
\BDBL {}Zhang, L.%
\end{APACrefauthors}%
\unskip\
\newblock
\APACrefYearMonthDay{2017}{}{}.
\newblock
{\BBOQ}\APACrefatitle {{The Subseasonal to Seasonal (S2S) Prediction Project
  Database}} {{The Subseasonal to Seasonal (S2S) Prediction Project
  Database}}.{\BBCQ}
\newblock
\APACjournalVolNumPages{Bulletin of the American Meteorological
  Society}{98}{1}{163--173}.
\newblock
\begin{APACrefDOI} \doi{10.1175/BAMS-D-16-0017.1} \end{APACrefDOI}
\PrintBackRefs{\CurrentBib}

\bibitem [\protect \citeauthoryear {%
Weigel%
, Baggenstos%
, Liniger%
, Vitart%
\BCBL {}\ \BBA {} Appenzeller%
}{%
Weigel%
\ \protect \BOthers {.}}{%
{\protect \APACyear {2008}}%
}]{%
Weigel2008}
\APACinsertmetastar {%
Weigel2008}%
\begin{APACrefauthors}%
Weigel, A\BPBI P.%
, Baggenstos, D.%
, Liniger, M\BPBI A.%
, Vitart, F.%
\BCBL {}\ \BBA {} Appenzeller, C.%
\end{APACrefauthors}%
\unskip\
\newblock
\APACrefYearMonthDay{2008}{dec}{}.
\newblock
{\BBOQ}\APACrefatitle {{Probabilistic Verification of Monthly Temperature
  Forecasts}} {{Probabilistic Verification of Monthly Temperature
  Forecasts}}.{\BBCQ}
\newblock
\APACjournalVolNumPages{Monthly Weather Review}{136}{12}{5162--5182}.
\newblock
\begin{APACrefURL} \url{http://journals.ametsoc.org/doi/10.1175/2008MWR2551.1}
  \end{APACrefURL}
\newblock
\begin{APACrefDOI} \doi{10.1175/2008MWR2551.1} \end{APACrefDOI}
\PrintBackRefs{\CurrentBib}

\bibitem [\protect \citeauthoryear {%
Weigel%
, Liniger%
\BCBL {}\ \BBA {} Appenzeller%
}{%
Weigel%
\ \protect \BOthers {.}}{%
{\protect \APACyear {2007}}%
}]{%
Weigel2007}
\APACinsertmetastar {%
Weigel2007}%
\begin{APACrefauthors}%
Weigel, A\BPBI P.%
, Liniger, M\BPBI A.%
\BCBL {}\ \BBA {} Appenzeller, C.%
\end{APACrefauthors}%
\unskip\
\newblock
\APACrefYearMonthDay{2007}{jan}{}.
\newblock
{\BBOQ}\APACrefatitle {{The discrete Brier and ranked probability skill
  scores}} {{The discrete Brier and ranked probability skill scores}}.{\BBCQ}
\newblock
\APACjournalVolNumPages{Monthly Weather Review}{135}{1}{118--124}.
\newblock
\begin{APACrefURL} \url{http://journals.ametsoc.org/doi/10.1175/MWR3280.1}
  \end{APACrefURL}
\newblock
\begin{APACrefDOI} \doi{10.1175/MWR3280.1} \end{APACrefDOI}
\PrintBackRefs{\CurrentBib}

\bibitem [\protect \citeauthoryear {%
Weyn%
, Durran%
\BCBL {}\ \BBA {} Caruana%
}{%
Weyn%
\ \protect \BOthers {.}}{%
{\protect \APACyear {2019}}%
}]{%
Weyn2019}
\APACinsertmetastar {%
Weyn2019}%
\begin{APACrefauthors}%
Weyn, J\BPBI A.%
, Durran, D\BPBI R.%
\BCBL {}\ \BBA {} Caruana, R.%
\end{APACrefauthors}%
\unskip\
\newblock
\APACrefYearMonthDay{2019}{}{}.
\newblock
{\BBOQ}\APACrefatitle {{Can Machines Learn to Predict Weather? Using Deep
  Learning to Predict Gridded 500‐hPa Geopotential Height From Historical
  Weather Data}} {{Can Machines Learn to Predict Weather? Using Deep Learning
  to Predict Gridded 500‐hPa Geopotential Height From Historical Weather
  Data}}.{\BBCQ}
\newblock
\APACjournalVolNumPages{Journal of Advances in Modeling Earth
  Systems}{11}{8}{2680--2693}.
\newblock
\begin{APACrefDOI} \doi{10.1029/2019ms001705} \end{APACrefDOI}
\PrintBackRefs{\CurrentBib}

\bibitem [\protect \citeauthoryear {%
Weyn%
, Durran%
\BCBL {}\ \BBA {} Caruana%
}{%
Weyn%
\ \protect \BOthers {.}}{%
{\protect \APACyear {2020}}%
}]{%
Weyn2020}
\APACinsertmetastar {%
Weyn2020}%
\begin{APACrefauthors}%
Weyn, J\BPBI A.%
, Durran, D\BPBI R.%
\BCBL {}\ \BBA {} Caruana, R.%
\end{APACrefauthors}%
\unskip\
\newblock
\APACrefYearMonthDay{2020}{}{}.
\newblock
{\BBOQ}\APACrefatitle {{Improving data-driven global weather prediction using
  deep convolutional neural networks on a cubed sphere}} {{Improving
  data-driven global weather prediction using deep convolutional neural
  networks on a cubed sphere}}.{\BBCQ}
\newblock
\APACjournalVolNumPages{Journal of Advances in Modeling Earth
  Systems}{12}{9}{e2020MS002109}.
\newblock
\begin{APACrefDOI} \doi{10.1029/2020MS002109} \end{APACrefDOI}
\PrintBackRefs{\CurrentBib}

\bibitem [\protect \citeauthoryear {%
White%
\ \protect \BOthers {.}}{%
White%
\ \protect \BOthers {.}}{%
{\protect \APACyear {2017}}%
}]{%
White2017}
\APACinsertmetastar {%
White2017}%
\begin{APACrefauthors}%
White, C\BPBI J.%
, Carlsen, H.%
, Robertson, A\BPBI W.%
, Klein, R\BPBI J\BPBI T.%
, Lazo, J\BPBI K.%
, Kumar, A.%
\BDBL {}Zebiak, S\BPBI E.%
\end{APACrefauthors}%
\unskip\
\newblock
\APACrefYearMonthDay{2017}{}{}.
\newblock
{\BBOQ}\APACrefatitle {{Potential applications of subseasonal-to-seasonal (S2S)
  predictions}} {{Potential applications of subseasonal-to-seasonal (S2S)
  predictions}}.{\BBCQ}
\newblock
\APACjournalVolNumPages{Meteorological Applications}{24}{3}{315--325}.
\newblock
\begin{APACrefDOI} \doi{10.1002/met.1654} \end{APACrefDOI}
\PrintBackRefs{\CurrentBib}

\end{thebibliography}

%
%
%
%
%

\end{document}